\documentclass{article} 
\usepackage{iclr2026_conference,times}
\usepackage{subcaption}
\usepackage{graphicx}
\usepackage{booktabs}
\usepackage{multirow}
\usepackage{xcolor}
\usepackage{float}

\usepackage{amsmath,amsfonts,bm}

\def\eqref#1{equation~\ref{#1}}

\def\1{\bm{1}}

\DeclareMathAlphabet{\mathsfit}{\encodingdefault}{\sfdefault}{m}{sl}
\SetMathAlphabet{\mathsfit}{bold}{\encodingdefault}{\sfdefault}{bx}{n}

\usepackage{hyperref}
\usepackage{url}
\usepackage{soul}
\usepackage{amssymb}
\usepackage{tikz}
\usepackage{pifont}
\usetikzlibrary{arrows.meta,calc,backgrounds}
\definecolor{hkHyper}{HTML}{E07A3F}     
\definecolor{hkHyperBg}{HTML}{FDF1E8}
\definecolor{hkFrozen}{HTML}{3D6FB6}    
\definecolor{hkFrozenBg}{HTML}{EAF1FA}
\definecolor{hkLatent}{HTML}{5B8C5A}
\definecolor{hkLatentBg}{HTML}{EDF5EC}
\definecolor{hkLoss}{HTML}{C0392B}
\definecolor{hkInk}{HTML}{2B2B2B}
\definecolor{hkNote}{HTML}{5A5A5A}
\newcommand{\hkfrz}{{\color{hkFrozen}\raisebox{0.05ex}{\ding{100}}}}
\newcommand{\hkm}[1]{{\normalsize$#1$}}   
\newcommand{\hkchunks}[2]{%
  \foreach \i in {-1,0,1}{\foreach \j in {-0.5,0.5}{%
    \node[chunk] at ($(#1,#2)+(\i*0.31,\j*0.31)$) {};}}}

\title{HyperKKL: Enabling Non-Autonomous State Estimation through Dynamic Weight Conditioning}

\author{Yahia Salaheldin Shaaban, Salem Lahlou \\
MBZUAI, Abu Dhabi, UAE \\
\texttt{\{Yahia.Shaaban,Salem.Lahlo\}@mbzuai.ac.ae} \\
\And
Abdelrahman Sayed Sayed \\
Univ Gustave Eiffel, COSYS-ESTAS, F-59657 \\
Villeneuve d’Ascq, France\\
\texttt{abdelrahman.ibrahim@univ-eiffel.fr} \\
}

\iclrfinalcopy 

\begin{document}

\maketitle

\begin{abstract}

This paper proposes HyperKKL, a novel learning approach for designing Kazantzis-Kravaris/Luenberger (KKL) observers for non-autonomous nonlinear systems. While KKL observers offer a rigorous theoretical framework by immersing nonlinear dynamics into a stable linear latent space, its practical realization relies on solving Partial Differential Equations (PDE) that are analytically intractable. Current existing learning-based approximations of the KKL observer are mostly designed for autonomous systems, failing to generalize to driven dynamics without expensive retraining or online gradient updates. HyperKKL addresses this by employing a hypernetwork architecture that encodes the exogenous input signal to instantaneously generate the parameters of the KKL observer, effectively learning a family of immersion maps parameterized by the external drive. We rigorously evaluate this approach against a curriculum learning strategy that attempts to generalize from autonomous regimes via training heuristics alone. The novel approach is illustrated on four numerical simulations in benchmark examples including the Duffing, Van der Pol, Lorenz, and Rössler systems.



\end{abstract}

\section{Introduction}
State estimation consider the reconstruction of the full internal state of a dynamical system from partial measurements is a foundational problem in science and engineering. In domains where dynamical models are used, from robotic control \citep{kang2025icra_mhe_grf_state, nisar2019vimo} to physiological monitoring \citep{hussain2021neural}, only a fraction of the state variables can be directly measured, and the rest must be inferred. The difficulty of this inverse problem scales with the complexity of the underlying dynamics: as systems become more nonlinear, higher-dimensional, and subject to external forcing. As a result, principled estimation methods become increasingly necessary, yet harder to design. A dynamical system is said to be \textit{autonomous} when its state evolution is governed entirely by its current state, without an external driving signal. However, real-world systems are almost never autonomous. Robotic platforms receive motor commands, biological systems respond to external stimuli, and industrial processes are subject to time-varying disturbances \citep{aastrom2021feedback}. These exogenous inputs fundamentally alter the state evolution and render such systems \textit{non-autonomous}. The interplay between these exogenous inputs and inherent physical nonlinearities yields \textit{non-autonomous nonlinear systems}, a regime that is ubiquitous in real-world applications yet presents the most challenges for observer synthesis.

Linear observers such as the Kalman filter \citep{kalman1960} and the Luenberger observer \citep{luenberger1964} are well studied, but they cannot capture the complexity of such systems. Nonlinear extensions, including the extended Kalman Filter and high-gain observers, exist but carry significant limitations as the former provides only local convergence guarantees \cite{julier1997new}, while the latter exhibits poor transient behavior and high noise sensitivity \citep{khalil2014high}. Kazantzis-Kravaris/Luenberger (KKL), another class of observers, offers a robust alternative \citep{kazantzis1998nonlinear,andrieu2006existence}. The core idea is to find a higher dimensional transformation that immerses the nonlinear dynamics into a space where the observer dynamics is linear and stable. Under the  observability condition known as backward distinguishability, this transformation is guaranteed to exist, allowing the state estimation to converge from any initial condition 
\citep{andrieu2006existence,bernard2024reconstructing}.

The practical bottleneck of KKL observers lies in their numerical realization. The transformation map satisfies a partial differential equation (PDE) that is analytically intractable for general nonlinear systems, and its left-inverse (map to original coordinate) is equally challenging to obtain  \citep{niazi_learning-based_2023}. Recent learning-based approaches have made significant progress by training neural networks to approximate these maps, either through supervised regression on simulated trajectories \citep{janny2021deepkkl}, through unsupervised learning objectives \citep{marani2025unsupervised}, through physics-informed (PINN) losses that explicitly encode the PDE constraint \citep{niazi2025kkl} or neural ordinary differential equations \citep{miao2023learning}.

A critical limitation of current learning-based KKL methods is that they are primarily designed for autonomous systems \citep{buisson2023towards}. Although the theoretical extension to non-autonomous settings was studied by \citet{bernard2019luenberger}, who showed that the KKL framework can accommodate exogenous inputs either by making the transformation maps input-dependent or by augmenting the observer dynamics with an additional injection term. However, no learning-based method has implemented such strategies in practice. On the other hand, meta-learning approaches, such as agnostic meta-learning, has been explored to adapt KKL observers to varying conditions \citep{trommer2023adaptive}, but this approach requires online gradient updates at test time and has only been validated on autonomous systems with parameter variations rather than exogenous inputs. Extending learning-based KKL observers to handle families of exogenous inputs without retraining remains an open problem, as explicitly noted by \citet{buisson2023towards} and \citet{niazi2025kkl}.

In this paper, we propose \textbf{HyperKKL}, a framework for extending KKL observers to \textbf{non-autonomous} systems, and for separating improvements due to \textbf{architecture} from those due to \textbf{training}. We study two complementary directions.

\begin{itemize}
    \item \textbf{Architecture} We employ hypernetwork conditioning \citep{ha2016hypernetworks}. A secondary network encodes the exogenous input signal and outputs the observer parameters. This yields an input-adaptive observer that can adjust at inference time without retraining or online gradient updates.
    \item \textbf{Training} We evaluate whether a \textbf{fixed-parameter KKL} observer can generalize to non-autonomous dynamics through training alone. We initialize from an observer pretrained in the autonomous setting and fine-tune on non-autonomous trajectories using \textbf{curriculum learning} that increases input complexity from simple signals to multi-frequency mixtures.
\end{itemize}

\textbf{Contributions.} (i) We introduce a \textbf{hypernetwork-conditioned KKL} observer that maps input signals to observer parameters, enabling adaptation to varying input conditions without retraining. (ii) We provide a controlled comparison against a training-only baseline and find that \textbf{curriculum-based transfer} from autonomous pretraining can \textbf{degrade performance} in some regimes. (iii) We present a \textbf{systematic empirical evaluation} in four nonlinear systems, \textbf{Duffing}, \textbf{Van der Pol}, \textbf{Lorenz}, and \textbf{R\"ossler}, comparing hypernetwork variants, curriculum schedules, and input encodings under various conditions.

\section{Preliminaries and Background}

\subsection{KKL Observer Theory for Non Linear Systems}
\label{subsection:KKL_Observer_Theory_for Autonomous_Systems}
\subsubsection{Autonmous systems formulation}
KKL observer relies on the immersion of a nonlinear system into a higher-dimensional linear system \citep{luenberger1964,luenberger1966,luenberger1971}. We consider an autonomous nonlinear systems of the form:
\begin{equation}
\begin{cases}
\dot{x} &= f(x), \\
y     &= h(x),
\end{cases}
\label{eq:KKL-autonomous_1}
\end{equation}
where $x \in \mathcal{X} \subset \mathbb{R}^{n_x}$ is the state at time
$t \in \mathbb{R}_{\geq0}$, $x_0 \in \mathcal{X}$ is the unknown initial
condition, and $y \in \mathbb{R}^{n_y}$ is the measured output. The
functions $f: \mathcal{X} \rightarrow \mathbb{R}^{n_x}$ and
$h: \mathcal{X} \rightarrow \mathbb{R}^{n_y}$ are assumed smooth.

The KKL framework seeks an injective mapping
$\mathcal{T}:\mathcal{X}\to\mathbb{R}^{n_z}$ (with
$n_z = n_y(2n_x + 1)$) that lifts the state to a higher-dimensional latent
space $z = \mathcal{T}(x)$ governed by linear dynamics. Specifically,
$\mathcal{T}$ must satisfy the PDE:
\begin{equation}
    \frac{\partial \mathcal{T}}{\partial x}(x)\, f(x)
    = A\,\mathcal{T}(x) + B\,h(x),
    \label{eq:KKL_autonomous_PDE}
\end{equation}
where $A \in \mathbb{R}^{n_z \times n_z}$ is Hurwitz,
$B \in \mathbb{R}^{n_z \times n_y}$, and $(A, B)$ is
controllable. This ensures that $z(t)$ evolves as
$\dot{z} = Az + Bh(x)$, and state estimation reduces to simulating a
linear observer:
\begin{equation}
    \dot{\hat{z}}(t) = A\hat{z}(t) + B\,y(t), \quad \hat{z}(0) = \hat{z}_0.
    \label{eq:KKL_observer_latent}
\end{equation}
Since $A$ is Hurwitz, the latent error
$e(t) = z(t) - \hat{z}(t) \to 0$ asymptotically~\citep[Remark
4]{niazi2025kkl}. A left-inverse
$\mathcal{T}^*: \mathbb{R}^{n_z} \to \mathcal{X}$ then recovers the state
estimate $\hat{x}(t) = \mathcal{T}^*(\hat{z}(t))$, ensuring
$\|x(t) - \hat{x}(t)\| \to 0$ as $t \to \infty$.

The existence of such a $\mathcal{T}$ is guaranteed under mild conditions:
\citet{brivadis2023further} show that if the system
is \textit{backward distinguishable} and \textit{forward complete} on a
compact $\mathcal{X}$, then an injective $\mathcal{T}$ satisfying~\eqref{eq:KKL_autonomous_PDE} exists.
\subsubsection{Extension to Non-Autonomous Systems}
\label{subsection:KKL_non_autonomous}

We consider a non-autonomous nonlinear systems of the form:
\begin{equation}
\begin{cases}
\dot{x} &= f(x, u(t)), \\
y     &= h(x),
\end{cases}
\label{eq:KKL_non-autonomous_1}
\end{equation}
where $x \in \mathcal{X} \subset \mathbb{R}^{n_x}$ is the state, $u(t) \in \mathbb{R}^{m}$ is a known external input, and $y(t) \in \mathbb{R}^{n_y}$ is the output. Extending the KKL framework to this setting is non-trivial: unlike the autonomous case, the requisite transformation $\mathcal{T}$ generally becomes \emph{input-dependent} ($\mathcal{T}_u$), often in a causal but implicit manner that hinders practical implementation \citep{bernard2019luenberger}. To address this, two paradigms exist: the \emph{stationary} approach and the \emph{dynamic} approach.

\paragraph{Stationary Transformation Approach}
Restricting attention to control-affine systems $\dot{x} = f(x) + g(x)u$, this method seeks a time-invariant diffeomorphism $\mathcal{T}: \mathcal{X} \to \mathbb{R}^{n_z}$ that maps the dynamics to a linear latent form with input injection: $\dot{\hat{z}} = A\hat{z} + B y + \bar{\varphi}(\hat{z}) u$. 
The input injection $\bar{\varphi}$ is then defined to satisfy the equivariance condition $\bar{\varphi}(\mathcal{T}(x)) = \frac{\partial \mathcal{T}}{\partial x}(x)g(x)$ \citep{bernard2019luenberger}. While architecturally simple (requiring no auxiliary states), this method is theoretically brittle. It requires the system to be uniformly instantaneously observable and the drift dynamics to be strongly differentially observable of order $n_x$. Consequently, it often fails for complex oscillators where inputs induce non-linear phase shifts that a static geometric map cannot capture.
\paragraph{Dynamic Transformation Approach}
To overcome the rigidity of static maps, the dynamic approach defines the transformation as a time-varying function $\mathcal{T}(x, t)$, or equivalently $\mathcal{T}(x, \theta(t))$, where $\theta(t)$ represents the state of an auxiliary dynamic filter driven by the input $u$. This transformation solves the time-dependent PDE:
\begin{equation}
    \frac{\partial \mathcal{T}}{\partial x}(x,t) f(x, u(t)) + \frac{\partial \mathcal{T}}{\partial t}(x,t) = A \mathcal{T}(x,t) + B h(x).
    \label{eq:dynamic_pde}
\end{equation}
The key advantage here is universality: \citet{bernard2019luenberger} prove that such a transformation exists under the much milder condition of backward distinguishability. Crucially, the inclusion of the partial time derivative $\frac{\partial \mathcal{T}}{\partial t}$ allows the transformation to ``slide'' along the solution manifold, effectively compensating for input-induced phase shifts or velocity changes that static maps cannot capture. The trade-off is increased computational complexity, as the observer must now learn to implicitly solve a time-varying PDE, necessitating the use of hypernetworks or recurrent architectures (e.g., LSTMs) to capture the required input history.
\subsection{Hypernetworks for Conditional Dynamics}
\label{subsection:Hypernetworks_for_Conditional_Dynamics}

A hypernetwork~\citep{ha2016hypernetworks,chauhan2024brief} is a neural network that generates, or modulates, the parameters of a \emph{target} (or \emph{base}) network conditioned on a context variable $c$. Given a target network $g_\theta$ with parameters $\theta$, a hypernetwork $\mathcal{H_\phi}$ parameterized by $\phi$ produces context-dependent weights:
\begin{equation}
    \theta = \mathcal{H_\phi}(c),
    \label{eq:hypernetwork}
\end{equation}
enabling the target network's behavior to adapt as a function of $c$, rather than remaining fixed across all conditions. In dynamical-systems settings, this paradigm has been used to build \emph{conditional dynamics models}. CODA~\citep{kirchmeyer2022generalizing} conditions a learned dynamics model on environment-specific context vectors inferred from observed trajectories via a hypernetwork with a low-rank locality constraint. HyperPINN~\citep{belbuteperes2021hyperpinn} and HyPINO~\citep{bischof2025hypino} generate PINN weights conditioned on PDE parameters to solve parameterized families of differential equations without retraining.

Many observer constructions, including the KKL class, are classically formulated for \emph{autonomous} systems. In the \emph{non-autonomous} case, the system evolves under an exogenous input $u(t)$, and the observer must account for input-dependent behavior. A natural way to achieve this is to condition the model on a representation of the input signal, allowing a subset of the model parameters to vary with the input, rather than with explicit physical parameters alone. In this work, we adopt this perspective. The context variable $c$ encodes a representation of the exogenous input, and the hypernetwork modulates the KKL inverse map accordingly. The precise conditioning mechanism and architecture are detailed in Section~\ref{sec:methodology}.

\subsection{Curriculum Learning}
\label{subsection:Curriculum_Learning}

Curriculum learning~\citep{bengio2009curriculum} is a training strategy inspired by human cognitive development. Instead of presenting examples in random order, the training data are organized by increasing difficulty, exposing the model to easier instances before harder ones. \citet{bengio2009curriculum} show that this can be viewed as a form of \emph{continuation method}, a classical strategy in non-convex optimization where a smoothed objective is gradually deformed toward the target objective, yielding both faster convergence and better local minima.

In the \emph{offline} variant of curriculum learning, all training data are pre-generated and partitioned into difficulty levels $\{\mathcal{D}_k\}_{k=1}^{K}$ before training begins. A difficulty metric defines the ordering, and a curriculum schedule governs when the learner advances to harder data. The progression can be \emph{adaptive}. The training procedure moves to the next difficulty level when the loss on the current level plateaus, thereby aligning curriculum transitions with learning progress.

In the context of KKL observer design, the core challenge of extending from autonomous to non-autonomous systems lies in conditioning the mappings or the observer, in our case we choose to extend the training on the inverse map $\mathcal{T}^*$, which must now account for exogenous inputs. We use curriculum learning to train this inverse map in an autoencoder-like fashion, sequentially exposing it to non-autonomous signals of increasing complexity. Here, \emph{difficulty} is associated with the spectral complexity of the exogenous input. Signals with richer high-frequency content produce more challenging system responses and, consequently, harder reconstruction targets for the inverse map. The concrete signal generation process, the difficulty metric, the curriculum schedule, and the plateau-detection rule are specified in Section~\ref{sec:methodology}.

\section{Methodology}
\label{sec:methodology}

\subsection{Problem Formulation}
\label{subsec:problem_formulation}

We consider observer design for nonautonomous nonlinear systems as defined
in~\eqref{eq:KKL_non-autonomous_1}. Our goal is to estimate $x(t)$ from $y(t)$
and $u(t)$ by immersing the nonlinear dynamics into the linear
observer~\eqref{eq:KKL_observer_latent}. As established in
Section~\ref{subsection:KKL_non_autonomous}, the nonautonomous setting
requires an input dependent transformation $\mathcal{T}(x, t)$ satisfying the
time varying PDE~\eqref{eq:dynamic_pde}, restated here for convenience:
\begin{equation}
    \frac{\partial \mathcal{T}}{\partial x}(x,t)\, f(x, u(t))
    + \frac{\partial \mathcal{T}}{\partial t}(x,t)
    = A\,\mathcal{T}(x,t) + B\,h(x).
    \label{eq:method_pde}
\end{equation}
Compared to the autonomous PDE~\eqref{eq:KKL_autonomous_PDE}, the additional
term $\frac{\partial \mathcal{T}}{\partial t}$ couples the transformation to the
temporal evolution of the input. This is precisely what makes static
transformations insufficient for time varying inputs: unless $\mathcal{T}$
adapts over time, the PDE residual grows with the rate of change of $u$.

Since analytical solutions to~\eqref{eq:method_pde} are intractable for
general $f(x, u)$, we cast the observer design as a learning problem. We
approximate both $\mathcal{T}$ and its left inverse $\mathcal{T}^*$ using
neural networks whose parameters are dynamically generated by a
hypernetwork $\mathcal{H}_\psi$ (Section~\ref{subsection:Hypernetworks_for_Conditional_Dynamics})
conditioned on the input history $u_{[t-w,\, t]}$.

Given a dataset $\mathcal{D}$ of trajectories $\{(x^{(i)}, u^{(i)},
y^{(i)})\}$, the learning objective minimizes the expected violation of the
KKL conditions:
\begin{equation}
    \min_{\psi} \; \mathbb{E}_{(x, u) \sim \mathcal{D}} \Big[
    \underbrace{\bigl\| x - \hat{\mathcal{T}}^*\bigl(\hat{\mathcal{T}}(x;\,
    \theta_u),\, \phi_u\bigr) \bigr\|^2}_{\mathcal{L}_{\text{rec}}}
    + \lambda
    \underbrace{\left\| \frac{\partial \hat{\mathcal{T}}}{\partial x} f(x, u)
    + \frac{\Delta \hat{\mathcal{T}}}{\Delta t}
    - A\hat{\mathcal{T}} - B\,h(x) \right\|^2}_{\mathcal{L}_{\text{PDE}}}
    \Big],
    \label{eq:optimization_objective}
\end{equation}
where $\theta_u, \phi_u$ denote the input conditioned parameters of the
encoder and decoder respectively, $\frac{\Delta \hat{\mathcal{T}}}{\Delta t}$
is a finite difference approximation of the temporal derivative, detailed in
Section~\ref{subsec:training}, and $\lambda > 0$ balances reconstruction
accuracy against dynamic consistency.

\subsection{HyperKKL Architecture}
\label{subsec:hyperkkl_arch}

The core idea of HyperKKL is to extend the autonomous KKL observer to
nonautonomous systems by conditioning the transformation maps on the
exogenous input $u(t)$. We build upon the physics informed autoencoder
of~\citet{niazi2025kkl}, comprising an encoder $\hat{\mathcal{T}}_\theta$
(the lifting map) and a decoder $\hat{\mathcal{T}}^*_\phi$ (the left inverse
map), and propose two architectures that implement the stationary and
dynamic paradigms of Section~\ref{subsection:KKL_non_autonomous}.

\subsubsection{Static HyperKKL, Stationary Approach}
\label{subsubsec:static}

Following the stationary formulation
of~\citet{bernard2019luenberger}, the Static HyperKKL retains the
autonomous transformation $\mathcal{T}(x)$ and augments the observer
dynamics with a learned input injection term:
\begin{equation}
    \dot{\hat{z}} = A\hat{z} + B\,y + \bar{\varphi}(\hat{z},\, u;\, \xi),
    \label{eq:static_observer}
\end{equation}
where $\bar{\varphi}$ is a small MLP parameterized by $\xi$ that
approximates the theoretically required injection
$\bar{\varphi}(\mathcal{T}(x)) = \tfrac{\partial \mathcal{T}}{\partial x}(x)\,g(x)$
from the control affine formulation in
Section~\ref{subsection:KKL_non_autonomous}. An LSTM encoder processes a
sliding window $u_{[t-w,\, t]}$ and produces a context embedding, which is
concatenated with $\hat{z}$ as input to the injection MLP. The injection
network is trained to output zero when $u \equiv 0$, ensuring exact recovery
of the autonomous observer on unforced dynamics.

This approach avoids modifying the learned maps $\hat{\mathcal{T}}_\theta$
and $\hat{\mathcal{T}}^*_\phi$, and instead compensates for the input's
effect entirely through the observer dynamics. It is most effective when the
input acts as a bounded perturbation that does not fundamentally alter the
attractor geometry.

\subsubsection{Dynamic HyperKKL, Dynamic Approach}
\label{subsubsec:dynamic}

For systems where the input continuously reshapes the
attractor, requiring a genuinely time varying transformation
$\mathcal{T}(x, t)$ as in~\eqref{eq:dynamic_pde}, we employ a residual
hypernetwork $\mathcal{H}_\psi$
(Section~\ref{subsection:Hypernetworks_for_Conditional_Dynamics}) that
modulates the weights of both the encoder and decoder. The context variable
is the input history $c = u_{[t-w,\, t]}$, and the parameters are decomposed
as:
\begin{align}
    \theta_{\text{enc}}(t) &= \theta_{\text{enc}}^{\text{base}}
    + \Delta\theta_{\text{enc}}\bigl(u_{[t-w,\,t]}\bigr), \\
    \phi_{\text{dec}}(t) &= \phi_{\text{dec}}^{\text{base}}
    + \Delta\phi_{\text{dec}}\bigl(u_{[t-w,\,t]}\bigr),
\end{align}
where $\theta^{\text{base}}_{\text{enc}}$ and
$\phi^{\text{base}}_{\text{dec}}$ are the frozen weights obtained from Phase 1
(Section~\ref{subsubsec:two_phase}), and $\Delta\theta$, $\Delta\phi$ are
input dependent perturbations generated by
$\mathcal{H}_\psi$.

\begin{figure}[H]
    \centering
\begin{tikzpicture}[
  x=1cm, y=1cm,
  font=\normalsize, text=hkInk,
  >={Stealth[length=2.1mm,width=1.55mm]},
  flow/.style={->, draw=hkInk, line width=0.7pt, rounded corners=2pt},
  hflow/.style={->, draw=hkHyper!78!black, line width=0.7pt, rounded corners=2pt},
  bflow/.style={->, draw=hkFrozen, line width=0.7pt},
  blk/.style={draw, line width=0.7pt, rounded corners=3pt, align=center, inner sep=2pt},
  hblk/.style={blk, draw=hkHyper, fill=hkHyperBg, minimum height=0.8cm,
              text height=0.66em, text depth=0pt},
  note/.style={font=\small, text=hkNote, inner sep=1pt},
  sig/.style={inner sep=1.5pt},
  plus/.style={circle, draw=hkHyper!78!black, fill=white, line width=0.7pt,
               minimum size=3.8mm, inner sep=0pt},
  chunk/.style={draw=hkHyper!78!black, fill=hkHyper!35, line width=0.35pt,
                minimum size=2.6mm, inner sep=0pt},
]

\def\xP{1.40}    
\def\xS{3.225}   
\def\xE{4.85}    
\def\xL{7.87}    
\def\xD{10.89}   
\def\xO{12.68}   
\def\xW{0.10}    
\def\xX{14.00}   
\def\yR{3.41}    
\def\yB{4.23}    
\def\yG{2.265}   
\def\yA{1.32}    
\def\yY{-1.62}   
\def\tw{2.0}     
\def\thS{0.55}   
\def\thL{0.90}   
\def\thM{0.725}  

\node[blk, draw=hkInk!55, fill=white, minimum width=2.2cm, minimum height=1.05cm]
  (sys) at (\xP,0) {\hkm{\dot{x}=f(x,u)}\\[-1pt]\hkm{y=h(x)}};
\node[note] at (\xP,-0.80) {system};

\draw[draw=hkFrozen, fill=hkFrozenBg, line width=0.7pt, rounded corners=1.5pt]
  (\xE-0.5*\tw,\thS) -- (\xE+0.5*\tw,\thL) -- (\xE+0.5*\tw,-\thL) -- (\xE-0.5*\tw,-\thS) -- cycle;
\node[align=center, inner sep=0pt] at (\xE+0.05,0)
  {\hkm{\hat{\mathcal{T}}_{\theta}}\\[-1pt]\hkm{\theta^{\mathrm{base}}\!+\!\Delta\theta}};
\node[note] at (\xE,-1.13) {encoder \hkfrz};

\node[blk, draw=hkLatent, fill=hkLatentBg, minimum width=2.3cm, minimum height=0.85cm]
  (lat) at (\xL,0) {\hkm{\dot{z}=Az+By}};
\node[note] at (\xL,0.72) {linear latent observer};

\draw[draw=hkFrozen, fill=hkFrozenBg, line width=0.7pt, rounded corners=1.5pt]
  (\xD-0.5*\tw,\thL) -- (\xD+0.5*\tw,\thS) -- (\xD+0.5*\tw,-\thS) -- (\xD-0.5*\tw,-\thL) -- cycle;
\node[align=center, inner sep=0pt] at (\xD-0.05,0)
  {\hkm{\hat{\mathcal{T}}^{*}_{\phi}}\\[-1pt]\hkm{\phi^{\mathrm{base}}\!+\!\Delta\phi}};
\node[note] at (\xD,-1.13) {decoder \hkfrz};

\node[sig] (xhat) at (\xO,0) {\hkm{\hat{x}_t}};

\draw[flow] (sys.east) -- node[above, sig] {\hkm{x_t}} (\xE-0.5*\tw,0);
\draw[flow] (\xE+0.5*\tw,0) -- node[above, sig] {\hkm{z_t}} (lat.west);
\draw[flow] (lat.east) -- node[above, sig] {\hkm{\hat{z}_t}} (\xD-0.5*\tw,0);
\draw[flow] (\xD+0.5*\tw,0) -- (xhat.west);

\draw[flow] (2.10,-0.525) -- (2.10,\yY) -- (\xL,\yY) -- (lat.south);
\node[sig, anchor=west] at (\xL+0.06,-1.02) {\hkm{y_t}};

\node[draw=hkLoss, dashed, line width=0.6pt, rounded corners=3pt, text=hkLoss, inner sep=2.5pt]
  (loss) at (\xO,-1.76) {\hkm{\mathcal{L}_{\mathrm{rec}}+\lambda\,\mathcal{L}_{\mathrm{PDE}}}};
\draw[->, dashed, draw=hkLoss, line width=0.7pt] (xhat.south) -- (loss.north);

\draw[draw=hkNote!55, fill=white, line width=0.5pt, rounded corners=1.5pt] (\xP-0.70,\yR-0.35) rectangle (\xP+0.70,\yR+0.35);
\draw[draw=hkHyper!78!black, line width=0.8pt, line cap=round] plot[smooth] coordinates {
  (\xP-0.62,\yR+0.00) (\xP-0.45,\yR+0.19) (\xP-0.27,\yR+0.21) (\xP-0.12,\yR-0.04)
  (\xP+0.05,\yR-0.23) (\xP+0.23,\yR-0.10) (\xP+0.40,\yR+0.17) (\xP+0.53,\yR+0.12)
  (\xP+0.62,\yR-0.03)};
\node[sig] at (\xP,\yR+0.63) {\hkm{u_{[t-w,\,t]}}};

\node[hblk, minimum width=1.25cm] (lstm) at (\xS,\yR) {LSTM};
\node[hblk, minimum width=1.5cm]  (mlpE) at (\xE,\yR) {MLP\hkm{{}_{\mathrm{enc}}}};
\node[hblk, minimum width=1.5cm]  (mlpD) at (\xD,\yR) {MLP\hkm{{}_{\mathrm{dec}}}};

\draw[-{Stealth[length=1.6mm,width=1.25mm]}, draw=hkHyper!78!black, line width=0.6pt]
  ($(lstm.south)+(0.15,0)$) -- ++(0,-0.16) arc[start angle=0, end angle=-180, radius=0.15] -- ++(0,0.16);

\draw[hflow] (\xP+0.70,\yR) -- (lstm.west);
\draw[hflow] (lstm.north) -- (\xS,\yB) -- (\xD,\yB) -- (mlpD.north);
\draw[hflow] (\xE,\yB) -- (mlpE.north);
\fill[hkHyper!78!black] (\xE,\yB) circle (1.4pt);
\node[sig] at (0.5*\xS+0.5*\xE,\yB+0.24) {\hkm{h_t}};

\hkchunks{\xE}{\yG}
\hkchunks{\xD}{\yG}
\node[plus] (addE) at (\xE,\yA) {};
\node[plus] (addD) at (\xD,\yA) {};
\foreach \n in {addE,addD}{
  \draw[hkHyper!78!black, line width=0.7pt] ($(\n.center)+(-0.105,0)$) -- ($(\n.center)+(0.105,0)$);
  \draw[hkHyper!78!black, line width=0.7pt] ($(\n.center)+(0,-0.105)$) -- ($(\n.center)+(0,0.105)$);
}
\foreach \x/\m/\a in {\xE/mlpE/addE, \xD/mlpD/addD}{
  \draw[hflow] (\m.south) -- (\x,\yG+0.345);
  \draw[hflow] (\x,\yG-0.345) -- (\a.north);
  \draw[hflow] (\a.south) -- (\x,\thM);
}
\node[note, anchor=east] at (\xE-0.58,\yG) {chunked};
\node[sig,  anchor=west] at (\xE+0.56,\yG) {\hkm{\Delta\theta}};
\node[note, anchor=east] at (\xD-0.58,\yG) {chunked};
\node[sig,  anchor=west] at (\xD+0.56,\yG) {\hkm{\Delta\phi}};

\node[sig, anchor=east] (bE) at (\xE-0.66,\yA) {\hkm{\theta^{\mathrm{base}}}\,\hkfrz};
\node[sig, anchor=east] (bD) at (\xD-0.66,\yA) {\hkm{\phi^{\mathrm{base}}}\,\hkfrz};
\draw[bflow] (bE.east) -- (addE.west);
\draw[bflow] (bD.east) -- (addD.west);

\draw[flow] (\xP,\yR-0.35) -- (sys.north);
\node[sig, anchor=west] at (\xP+0.06,1.11) {\hkm{u_t}};

\begin{scope}[on background layer]
  \fill[hkHyperBg!55, rounded corners=6pt]  (\xW,1.81)  rectangle (\xX,4.89);
  \fill[hkFrozenBg!45, rounded corners=6pt] (\xW,-2.63) rectangle (\xX,1.69);
\end{scope}
\node[anchor=east, inner sep=0pt, font=\small\bfseries, text=hkHyper!72!black] at (\xX-0.17,4.59)
  {Input-conditioned hypernetwork \textmd{(trained in Phase 2)}};
\node[anchor=east, inner sep=0pt, font=\small\bfseries, text=hkFrozen] at (\xX-0.17,-2.34)
  {KKL observer \textmd{(base maps pre-trained in Phase 1, frozen \hkfrz)}};

\end{tikzpicture}
    \caption{\textbf{Dynamic HyperKKL Architecture (Phase~2).} The base
    encoder $\hat{\mathcal{T}}_{\theta^{\text{base}}}$ and decoder
    $\hat{\mathcal{T}}^*_{\phi^{\text{base}}}$, pre-trained on autonomous
    dynamics in Phase~1, are frozen. A shared LSTM encoder processes the
    input window $u_{[t-w,\,t]}$ and produces a hidden state $h_t$, which is
    passed to two separate MLP decoder heads that predict chunked weight
    perturbations $\Delta\theta_{\text{enc}}$ and
    $\Delta\phi_{\text{dec}}$. The perturbations are added to the frozen
    base weights, yielding input-conditioned maps
    $\hat{\mathcal{T}}(x;\,\theta^{\text{base}} + \Delta\theta)$ and
    $\hat{\mathcal{T}}^*(z;\,\phi^{\text{base}} + \Delta\phi)$.}
    \label{fig:architecture}
\end{figure}
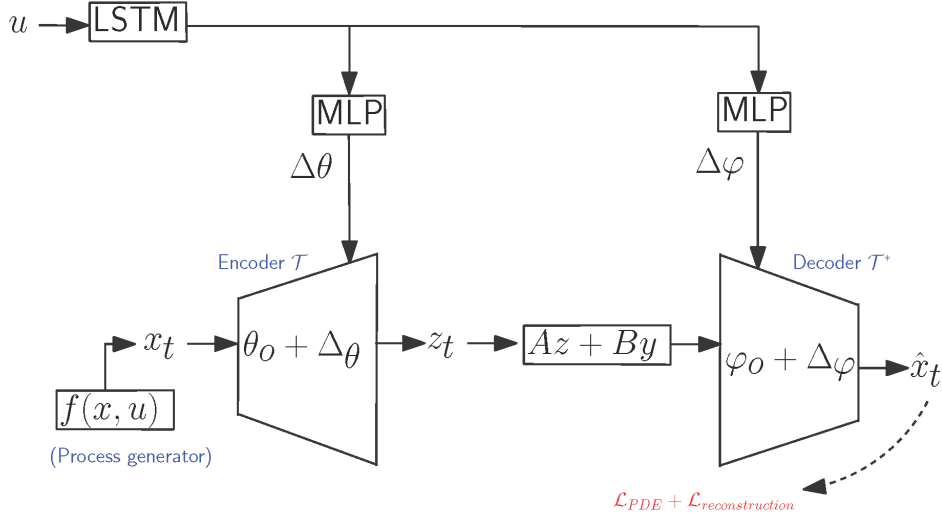

The hypernetwork consists of three components. First, a \textbf{shared LSTM
encoder} processes the input window $u_{[t-w,\, t]}$ and produces a hidden
state $h_t \in \mathbb{R}^{d_h}$ summarizing the input history.This hidden state is then passed to \textbf{two separate MLP decoder heads},
one for $\Delta\theta_{\text{enc}}$ and one for
$\Delta\phi_{\text{dec}}$. Since directly predicting the full weight
perturbation for each target layer would be prohibitively large, each
decoder head employs a \textbf{chunked prediction} strategy: the target
weight matrix $W \in \mathbb{R}^{m \times n}$ is partitioned into smaller
blocks, and the MLP predicts each chunk independently from the shared LSTM
embedding. This keeps the decoder output dimension manageable while
preserving full-rank expressivity within each chunk, avoiding the
representational bottleneck of low-rank factorizations (As illustrated in Figure~\ref{fig:architecture}).

This residual structure ensures that when $u \equiv 0$, the LSTM hidden
state produces $\Delta\theta = \Delta\phi = 0$, exactly recovering the
autonomous observer. The low rank constraint further acts as an implicit
regularizer, preventing the hypernetwork from overperturbing the
well trained base maps.

\subsection{Training Procedure}
\label{subsec:training}
\subsubsection{Two Phase Sequential Training}
\label{subsubsec:two_phase}

Following the sequential training paradigm of~\citet{niazi2025kkl} to avoid
gradient conflict between the encoder and decoder objectives, we adopt a
two phase procedure:

\textbf{Phase 1, Autonomous Pretraining.} We train the base encoder
$\hat{\mathcal{T}}_{\theta^{\text{base}}}$ and decoder
$\hat{\mathcal{T}}^*_{\phi^{\text{base}}}$ on unforced dynamics
($u \equiv 0$) using the physics informed loss
from~\citet{niazi2025kkl}, comprising a data fit term
$\|z - \hat{\mathcal{T}}(x)\|^2$ on labeled trajectory pairs and the
autonomous PDE residual~\eqref{eq:KKL_autonomous_PDE} evaluated on
collocation points. This produces maps that satisfy the autonomous KKL
conditions and establishes a high quality initialization. Upon completion,
$\theta^{\text{base}}$ and $\phi^{\text{base}}$ are frozen.

\textbf{Phase 2, Hypernetwork Training.} With the base maps fixed, we
train only the hypernetwork parameters $\psi$ on forced trajectories. The
loss~\eqref{eq:optimization_objective} is computed using the
input conditioned maps $\hat{\mathcal{T}}(x;\,\theta^{\text{base}} +
\Delta\theta)$ and $\hat{\mathcal{T}}^*(z;\,\phi^{\text{base}} +
\Delta\phi)$. The spatial gradient $\frac{\partial
\hat{\mathcal{T}}}{\partial x}$ is computed via automatic differentiation,
while the temporal derivative $\frac{\partial \mathcal{T}}{\partial t}$
in~\eqref{eq:method_pde} is approximated via finite differences over
consecutive input windows:
\begin{equation}
    \frac{\Delta \hat{\mathcal{T}}}{\Delta t}
    \approx \frac{
    \hat{\mathcal{T}}\bigl(x;\,\theta(u_{[t,\, t+\Delta t]})\bigr)
    - \hat{\mathcal{T}}\bigl(x;\,\theta(u_{[t-\Delta t,\, t]})\bigr)
    }{\Delta t}.
    \label{eq:finite_diff}
\end{equation}
That is, for the same state $x$, we evaluate the encoder under two adjacent
input windows and take their difference. This captures how the
transformation evolves as the input shifts, without requiring explicit
differentiation through the LSTM.

\subsubsection{Adaptive Curriculum Learning, Baseline}
\label{subsubsec:curriculum}

As a training strategy baseline, we evaluate the \textbf{Adaptive
Curriculum} approach described in
Section~\ref{subsection:Curriculum_Learning}. This method uses the same
static architecture as the Autonomous observer but trains
$\hat{\mathcal{T}}^*$ on nonautonomous data of progressively increasing
difficulty.

Training trajectories are generated offline and partitioned into difficulty
levels $\{\mathcal{D}_k\}_{k=1}^{K}$ ordered by spectral complexity: $k=1$
corresponds to constant inputs ($u = c$), $k=2$ to low frequency sinusoids,
and subsequent levels introduce higher frequency components and mixtures. The
difficulty metric is defined by the dominant frequency content and rate of
change $\|\dot{u}\|$ of each trajectory's input signal. Training proceeds on
level $\mathcal{D}_k$ until the loss plateaus, detected when the relative
improvement falls below a threshold $\epsilon$ over a patience window of $p$
epochs, at which point the scheduler advances to $\mathcal{D}_{k+1}$. This
tests whether data diversity alone, without architectural
changes, can bridge the gap to nonautonomous observation.

\section{Experimental Evaluation}
\label{sec:experiments}

We evaluate the proposed HyperKKL framework over four nonlinear benchmarks described in detail in Appendices \ref{apd:Duffing system}, \ref{apd:Van der Pol system}, \ref{apd:Rössler system}, and \ref{apd:Lorenz system}. Where they are classified into two low-dimensional oscillators (Duffing, Van der Pol), and two chaotic systems (Rossler, Lorenz). 

\subsection{Experimental Setup}

\textbf{Datasets and Tasks.} For each system, we generate synthetic trajectories via RK4 integration ($\Delta t = 0.05s$, horizon $T=50s$). Process and measurement noise are added ($\sigma=0.01$). We evaluate the generalization on four input regimes $u(t)$: (i) \textbf{Zero} (autonomous), (ii) \textbf{Constant} ($u \sim \mathcal{U}[-1, 1]$), (iii) \textbf{Sinusoid} (randomized $A, \omega, \phi$), and (iv) \textbf{Square Wave} (discontinuous jumps).

\textbf{Baselines.} We compare against two standard approaches:(i) \textbf{Autonomous KKL}, the observer trained solely on unforced dynamics ($u=0$), which quantifies the domain shift caused by external inputs; and (ii) \textbf{Adaptive Curriculum}, the same static inverse map $\mathcal{T}^*$
retrained on progressively complex input regimes (constant $\to$ sinusoidal $\to$ square wave),
testing whether data diversity alone can overcome architectural limitations.

\textbf{Our Framework (HyperKKL).} We propose two variants of the HyperKKL architecture:
\begin{itemize}
    \item \textbf{Static HyperKKL:} A stationary transformation $\mathcal{T}(x)$
    paired with a learned input injection $\bar{\varphi}(z, u)$ that conditions the
    observer dynamics on the instantaneous input. This implements the
    ``Stationary Approach'' (Sec.~\ref{subsection:KKL_non_autonomous}).
    \item \textbf{Dynamic HyperKKL:} The full time-varying transformation $\mathcal{T}(x, \theta(t))$ where parameters $\theta(t)$ are generated by an LSTM-based hypernetwork. This explicitly solves the dynamic PDE \eqref{eq:dynamic_pde} by adapting to the input history.
\end{itemize}

\textbf{Implementation Details \& Training Strategy.} All encoders are 3-layer MLPs (150 units for oscillators, 350 for chaotic systems). The Dynamic HyperKKL employs an LSTM hypernetwork (64 units, window $w=100$) predicting encoder weights via a low-rank decomposition (Rank: 32 for oscillators, 128 for chaotic systems). Following the physics-informed framework of \citet{niazi2025kkl}, we adopt a \textbf{sequential training} scheme to mitigate gradient conflict: we first \textbf{warm-start} the base encoder $\mathcal{T}_{\text{base}}$ on autonomous dynamics (Phase 1), then train the hypernetwork on forced dynamics (Phase 2). To prevent loss explosion on large-magnitude chaotic systems (e.g., Lorenz), we normalize the vector field term $f(x)$ in the PDE loss and apply \textbf{gradient clipping} (norm 1.0). We adopt the data set generation protocol and autonomous baselines from \citet{niazi2025kkl}, as they represent the current state-of-the-art for autonomous KKL.

\textbf{Evaluation Metrics.} We assess state estimation accuracy using the \textbf{Root Mean Squared Error (RMSE)} averaged over test trajectories generated with input parameters distinct from those in the training set, ensuring that we evaluate generalization to unseen forcing regimes. To account for the varying scales of different systems, we also report the \textbf{Symmetric Mean Absolute Percentage Error (SMAPE)}. Both metrics are computed on the steady-state response (ignoring the first 5\% of the trajectory) to focus on long-term tracking capability rather than initial transient convergence.

\subsection{Results and Analysis}
\label{sec:results_discussion}

Tables~\ref{tab:results_part1} and~\ref{tab:results_part2} report RMSE and SMAPE across four systems and four input conditions. We organize our analysis around three key findings.

\textbf{Hypernetwork Approaches Improve State Estimation on Oscillatory and Mildly Chaotic Systems.}
On the Duffing oscillator, the \textbf{Static HyperKKL} achieves the strongest results, reducing RMSE by up to $62\%$ relative to the Autonomous baseline under sinusoidal inputs ($0.26 \to 0.10$) and $48\%$ under square wave inputs ($0.33 \to 0.17$). This is consistent with the theoretical expectation: for low dimensional oscillators whose attractor geometry shifts smoothly with a slowly varying input, a \emph{stationary} transformation $\mathcal{T}(x, u)$ suffices to maintain injectivity. On the Van der Pol oscillator, both hypernetwork variants improve over the autonomous baseline under time varying inputs, with the \textbf{Dynamic HyperKKL} achieving the best RMSE on sinusoidal ($0.21$) and square wave ($0.22$) forcing. For the chaotic Rössler system, the Dynamic approach yields the lowest errors across all non zero input types (e.g., $1.48 \to 1.36$ under square wave input), confirming that temporal aggregation of input history becomes increasingly important as attractor complexity grows.

\textbf{Catastrophic Failure of Curriculum Learning and Static Conditioning on Chaotic Attractors.}
\textbf{Curriculum Learning} performs dramatically worse than even the naive Autonomous baseline on every system and input condition tested. For instance, on the Van der Pol oscillator RMSE increases from $0.15$ to $1.10$ under zero input, and from $0.25$ to $1.15$ under square wave input. On Lorenz, RMSE roughly doubles ($5.55 \to 11.6$). This demonstrates that the bottleneck is \emph{representational}, not educational: exposing a static architecture to progressively complex non autonomous instances cannot compensate for an inductive bias that is mathematically incapable of solving the required dynamic PDE. Similarly striking is the behavior of the \textbf{Static HyperKKL} on the Lorenz system, where conditioning on instantaneous input leads to catastrophic degradation (RMSE $\approx 16$ under all non zero inputs, versus $5.5$ for the Autonomous baseline). This confirms the theoretical analysis of \citet{bernard2019observer}: for highly sensitive chaotic systems, a static transformation $\mathcal{T}(x, u(t))$ is insufficient to ensure injectivity, and conditioning on the wrong information can be \emph{actively harmful}, distorting the learned immersion rather than refining it.

\textbf{Theoretical Consistency and Limitations on the Lorenz Attractor.}
Across all four systems, the hypernetwork methods correctly recover the autonomous baseline with zero input ($u = 0$), empirically validating the architectural constraint $\Delta\theta \to 0$ as $u \to 0$. However, the Lorenz system exposes a fundamental limitation: the \textbf{Autonomous} baseline achieves the best overall performance (RMSE $\approx 5.5$), and even the Dynamic HyperKKL, while dramatically outperforming Curriculum ($11.6 \to 6.67$) and Static ($16.2 \to 6.66$), incurs a modest degradation relative to the input agnostic observer ($5.55 \to 6.66$). We attribute this to the extreme sensitivity of the Lorenz attractor to perturbations: small errors in the hypernetwork's input conditioned weight modulation propagate exponentially along unstable manifolds, producing residual estimation drift that a conservative autonomous observer avoids by ignoring the input entirely. This suggests that for systems near the edge of observability, the additional representational capacity of hypernetworks must be paired with explicit stability guarantees or regularization to prevent the input conditioning pathway from introducing more noise than signal. Addressing this trade off, potentially through Lyapunov informed training constraints or adaptive gating mechanisms that attenuate modulation under high sensitivity, is an important direction for future work.

\begin{figure}[h]
    \centering
    \includegraphics[width=0.95\textwidth]{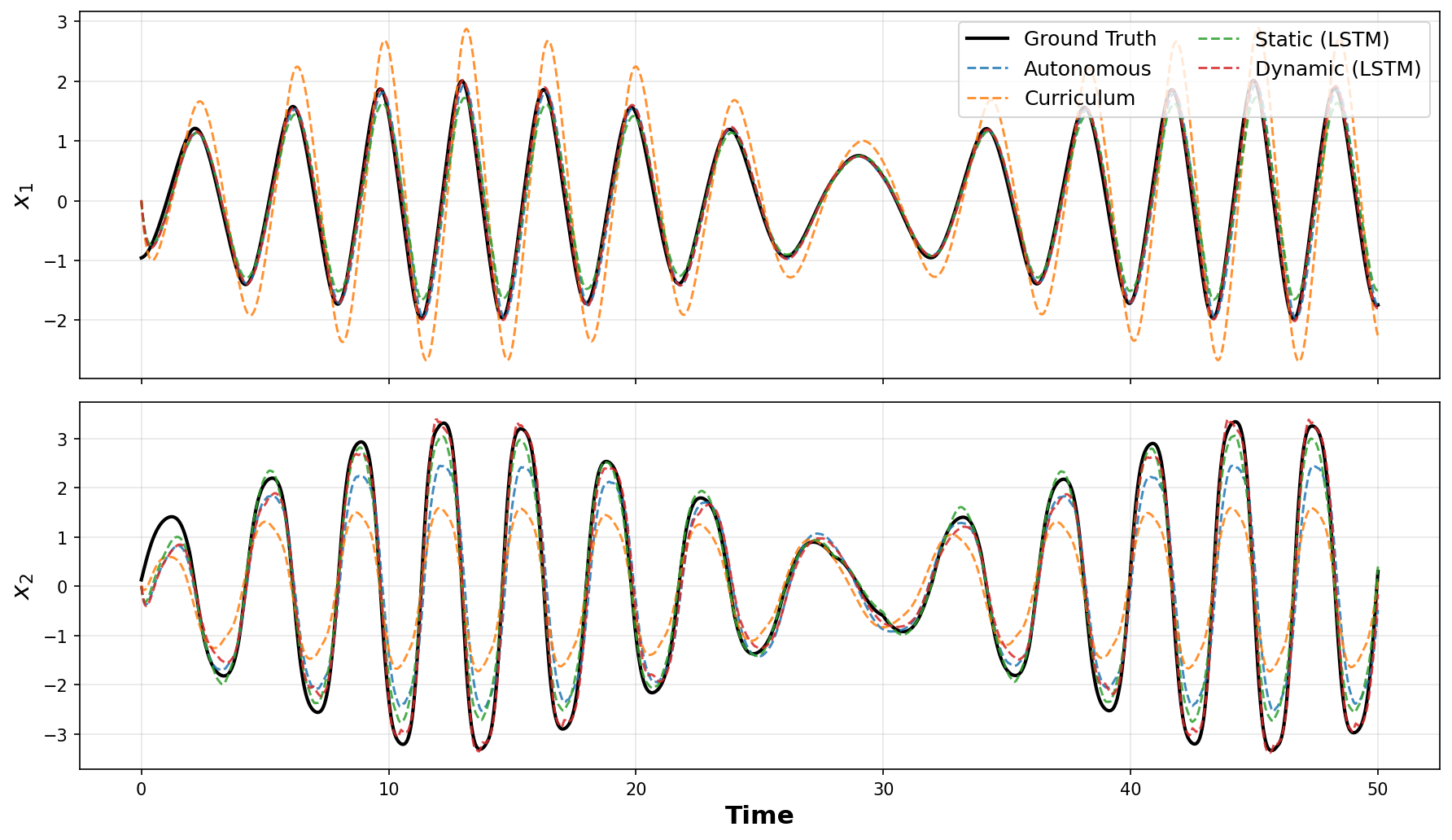}
    \caption{\textbf{Duffing (Square Input):} The Dynamic HyperKKL adapts to discontinuous square waves without the transient spikes seen in static methods.}
    \label{fig:duffing_square}
\end{figure}

Figure~\ref{fig:duffing_square} illustrates the qualitative behavior on the Duffing oscillator under discontinuous square wave forcing. The Dynamic HyperKKL tracks the ground truth smoothly across input transitions, whereas the Static method exhibits transient spikes at discontinuities, and the Autonomous and Curriculum baselines accumulate persistent phase drift.

\begin{table}[H]
\centering
\caption{State Estimation Performance (Duffing and Van der Pol): \textbf{RMSE} \textcolor{gray}{\small (SMAPE \%)}. Lower is better.}
\label{tab:results_part1}
\resizebox{\textwidth}{!}{%
\begin{tabular}{l|cccc|cccc}
\toprule
\multirow{2}{*}{\textbf{Method}} & \multicolumn{4}{c|}{\textbf{Duffing}} & \multicolumn{4}{c}{\textbf{Van der Pol}} \\
\cmidrule(lr){2-5} \cmidrule(lr){6-9}
& Zero & Const & Sin & Sqr & Zero & Const & Sin & Sqr \\
\midrule
Autonomous & 0.04 \textcolor{gray}{(5.6)} & 0.63 \textcolor{gray}{(66)} & 0.26 \textcolor{gray}{(26)} & 0.33 \textcolor{gray}{(31)} & 0.15 \textcolor{gray}{(7.0)} & 0.35 \textcolor{gray}{(23.1)} & 0.23 \textcolor{gray}{(9.8)} & 0.25 \textcolor{gray}{(10.5)} \\
Curriculum & 0.27 \textcolor{gray}{(33)} & 0.64 \textcolor{gray}{(63)} & 0.44 \textcolor{gray}{(41)} & 0.57 \textcolor{gray}{(46)} & 1.10 \textcolor{gray}{(51.4)} & 1.00 \textcolor{gray}{(54.3)} & 1.15 \textcolor{gray}{(51.5)} & 1.15 \textcolor{gray}{(51.7)} \\
\midrule
Static HyperKKL & \textbf{0.04} \textcolor{gray}{(5.6)} & \textbf{0.39}$\downarrow$ \textcolor{gray}{(38)} & \textbf{0.10}$\downarrow$ \textcolor{gray}{(9.3)} & \textbf{0.17}$\downarrow$ \textcolor{gray}{(14)} & 0.12$\downarrow$ \textcolor{gray}{(5.3)} & \textbf{0.26}$\downarrow$ \textcolor{gray}{(14.1)} & 0.24 \textcolor{gray}{(10.2)} & 0.25 \textcolor{gray}{(10.8)} \\
Dynamic HyperKKL & 0.08 \textcolor{gray}{(8.2)} & 0.56$\downarrow$ \textcolor{gray}{(62)} & 0.24$\downarrow$ \textcolor{gray}{(25)} & 0.27$\downarrow$ \textcolor{gray}{(28)} & \textbf{0.12}$\downarrow$ \textcolor{gray}{(5.0)} & 0.38 \textcolor{gray}{(25.6)} & \textbf{0.21}$\downarrow$ \textcolor{gray}{(8.6)} & \textbf{0.22}$\downarrow$ \textcolor{gray}{(9.1)} \\
\bottomrule
\end{tabular}%
}
\end{table}

\begin{table}[H]
\centering
\caption{State Estimation Performance (Rossler and Lorenz): \textbf{RMSE} \textcolor{gray}{\small (SMAPE \%)}. Lower is better.}
\label{tab:results_part2}
\resizebox{\textwidth}{!}{%
\begin{tabular}{l|cccc|cccc}
\toprule
\multirow{2}{*}{\textbf{Method}} & \multicolumn{4}{c|}{\textbf{Rossler (Chaotic)}} & \multicolumn{4}{c}{\textbf{Lorenz (Chaotic)}} \\
\cmidrule(lr){2-5} \cmidrule(lr){6-9}
& Zero & Const & Sin & Sqr & Zero & Const & Sin & Sqr \\
\midrule
Autonomous & 1.14 \textcolor{gray}{(6.7)} & 1.75 \textcolor{gray}{(8.6)} & 1.47 \textcolor{gray}{(7.6)} & 1.48 \textcolor{gray}{(8.3)} & \textbf{5.56} \textcolor{gray}{(18)} & \textbf{5.50} \textcolor{gray}{(18)} & \textbf{5.58} \textcolor{gray}{(18)} & \textbf{5.55} \textcolor{gray}{(18)} \\
Curriculum & 5.58 \textcolor{gray}{(35)} & 5.55 \textcolor{gray}{(35)} & 5.94 \textcolor{gray}{(37)} & 5.61 \textcolor{gray}{(38)} & 11.5 \textcolor{gray}{(41)} & 11.4 \textcolor{gray}{(41)} & 11.6 \textcolor{gray}{(42)} & 11.6 \textcolor{gray}{(42)} \\
\midrule
Static HyperKKL & 1.14 \textcolor{gray}{(6.7)} & \textbf{1.50}$\downarrow$ \textcolor{gray}{(9.5)} & 1.70 \textcolor{gray}{(10)} & 1.75 \textcolor{gray}{(12)} & \textbf{5.56} \textcolor{gray}{(18)} & 16.0 \textcolor{gray}{(51)} & 16.3 \textcolor{gray}{(52)} & 16.2 \textcolor{gray}{(51)} \\
Dynamic HyperKKL & \textbf{1.01}$\downarrow$ \textcolor{gray}{(5.1)} & 1.57$\downarrow$ \textcolor{gray}{(7.7)} & \textbf{1.38}$\downarrow$ \textcolor{gray}{(6.0)} & \textbf{1.36}$\downarrow$ \textcolor{gray}{(6.9)} & 6.67 \textcolor{gray}{(22)} & 6.64 \textcolor{gray}{(22)} & 6.67 \textcolor{gray}{(22)} & 6.66 \textcolor{gray}{(22)} \\
\bottomrule
\end{tabular}%
}
\end{table}

\section{Conclusion and Future Work}
\label{sec:conclusion}

We introduce \textbf{HyperKKL}, a framework that extends physics-informed KKL observer design to non autonomous nonlinear systems by conditioning observer parameters on input history via hypernetworks. Our experiments reveal that hypernetwork based observers substantially reduce estimation error on oscillatory and mildly chaotic systems under time varying forcing, while the catastrophic failure of Curriculum Learning across all benchmarks confirmsthe challenge is \emph{representational}, not a lack of training on exogenous input dynamics. However, on the highly sensitive Lorenz attractor, the Autonomous observer remains strongest, exposing a fundamental tension: input conditioning adds capacity that can improve tracking, but also introduces a modulation pathway through which errors propagate along unstable manifolds. The divergent failure modes, with Curriculum failing everywhere and Static HyperKKL collapsing on Lorenz, underscore that naive input conditioning can be worse than ignoring the input entirely, motivating architectures grounded in the structure of the underlying PDE.

\textbf{Future Work.} The Lorenz failure mode suggests that hypernetwork modulation requires explicit stability constraints, such as Lyapunov informed regularization or adaptive gating that attenuates weight perturbations in high sensitivity regions. Disentangling phase modulation from amplitude scaling within the architecture may further improve robustness. Finally, extending to partially known dynamics would bridge the gap between the idealized benchmarks studied here and grey box industrial settings with model uncertainty and sparse sensor data.
\clearpage
\bibliography{iclr2026_conference}
\bibliographystyle{iclr2026_conference}

\clearpage
\appendix

\section{System description}\label{apd:system}

\textbf{Experiment Settings:} All experiments herein are run on a single server equipped with NVIDIA RTX A6000 (50 GB Vram), 48 CPU cores, and 256 GB RAM with PyTorch library V2.9.


\subsection{Reverse Duffing Oscillator}\label{apd:Duffing system}

The Reverse Duffing oscillator is a variant of the classical oscillator studied by Georg Duffing to model structural instability and chaotic behavior in mechanical systems \citep{Korsch1999}. It features a cubic nonlinearity in the stiffness term, making it a standard benchmark for evaluating the observer performance in the presence of strong, non-Lipschitz nonlinearities. We consider the continuous-time dynamics given by:
$$\begin{cases}
\dot{x}_1 = x_2^3, \\
\dot{x}_2 = -x_1, \\
y = x_1,
\end{cases}$$
where $x = [x_1, x_2]^\top \in \mathbb{R}^2$ is the state vector and $y \in \mathbb{R}$ is the measured output. This system is particularly challenging due to the aggressive cubic growth of the $x_2^3$ term, testing the observer's ability to handle fast dynamics far from the equilibrium.

\subsection{Van der Pol Oscillator}\label{apd:Van der Pol system}

The Van der Pol oscillator is a 2-dimensional limit cycle oscillator originally by Van der Pol to model electrical circuits with vacuum tubes \citep{Ginoux2017}. It is characterized by a nonlinear damping that dissipates energy at high amplitudes but generates energy at low amplitudes, resulting in a stable limit cycle, making it valuable for studying self-sustained oscillations in biological and physical systems. We consider the parameterized dynamics with $\mu = 3$:
$$\begin{cases}
\dot{x}_1 = x_2, \\
\dot{x}_2 = \mu(1 - x_1^2)x_2 - x_1, \\
y = x_1,
\end{cases}$$
where $x = [x_1, x_2]^\top \in \mathbb{R}^2$ is the state vector, $\mu > 0$ determines the nonlinearity and damping strength, and $y \in \mathbb{R}$ is the measured output. The parameter $\mu$ significantly alters the stiffness of the system, making it ideal for testing the hypernetwork ability to handle the varying physical parameters.

\subsection{Rössler Attractor}\label{apd:Rössler system}

The Rössler attractor is a 3-dimensional chaotic system introduced by \citet{rossler1976equation}, designed to be one of the simplest continuous-time systems capable of exhibiting chaotic behavior. It is smoother than the Lorenz attractor \ref{apd:Lorenz system} but still produces complex, fractal-like trajectories, serving as a robust testbed for reconstructing higher-dimensional chaotic states from lower-dimensional measurements. The attractor dynamics are defined as:
$$\begin{cases}
\dot{x}_1 = -x_2 - x_3, \\
\dot{x}_2 = x_1 + ax_2, \\
\dot{x}_3 = b + x_3(x_1 - c), \\
y = x_2,
\end{cases}$$
where $x = [x_1, x_2, x_3]^\top \in \mathbb{R}^3$ is the state vector. We utilize a standard set of chaotic parameters $a = 0.1$, $b = 0.1$, and $c = 14$. The system output is $y = x_2$, requiring the observer to reconstruct the full 3-dimensional chaotic state from a single measurement channel.

\subsection{Lorenz System}\label{apd:Lorenz system}

The Lorenz attractor is a seminal 3-dimensional system originally derived by \citet{DeterministicNonperiodicFlow} for atmospheric convection. It is arguably the most famous example of deterministic chaos, mainly known for its "butterfly effect" where sensitive dependence on initial conditions makes long-term prediction impossible without accurate state estimation. The system dynamics are given by:
$$\begin{cases}
\dot{x}_1 = p(x_2 - x_1) \\
\dot{x}_2 = x_1(q - x_3) - x_2 \\
\dot{x}_3 = x_1 x_2 - r x_3 \\
y = x_2
\end{cases}$$
where $x = [x_1, x_2, x_3]^\top \in \mathbb{R}^3$ is the state vector. We use the classic parameter values $p = 10$, $q = 28$, and $r = 8/3$.

\section{Connection to Non-Autonomous KKL Theory}\label{apd:nonautonomous_KKL_theory}

In this appendix, we establish the theoretical foundations of the proposed
HyperKKL framework. We first visit the KKL observer theory for autonomous
systems and its extension to non-autonomous systems via physics-informed
learning.

\subsection{KKL Observer Theory for Non-Autonomous Systems}\label{apd:KKL_Observer_Theory_for_Non-Autonomous_Systems}

KKL observer relies on the immersion of a nonlinear system into a
higher-dimensional linear
system~\citep{luenberger1964,luenberger1966,luenberger1971}. We consider
non-autonomous nonlinear systems of the
form~\eqref{eq:KKL_non-autonomous_1}, restated here:
\begin{equation}
\begin{cases}
\dot{x}(t) &= f(x(t), u(t)), \\
y(t)       &= h(x(t))
\end{cases}
\label{apd:eq_system}
\end{equation}
where $x(t) \in \mathcal{X} \subset \mathbb{R}^{n_x}$ is the state at time
$t \in \mathbb{R}_{\geq0}$, $x_0 \in \mathcal{X}$ is the unknown initial
condition, $u(t) \in \mathbb{R}^{m}$ is the external control input, and
$y(t) \in \mathbb{R}^{n_y}$ is the measured output. The functions
$f: \mathcal{X} \times \mathbb{R}^{m} \rightarrow \mathbb{R}^{n_x}$ and
$h: \mathcal{X} \rightarrow \mathbb{R}^{n_y}$ are assumed to be smooth. The
KKL framework involves designing an observer of the following form:
\begin{equation}
\begin{cases}
\dot{\hat{z}}(t) &= \Phi(\hat{z}(t), y(t), u(t)), \\
\hat{x}(t)       &= \Psi(\hat{z}(t), y(t))
\end{cases}
\label{apd:eq_observer_general}
\end{equation}
where $\hat{z}(t) \in \mathbb{R}^{n_{z}}$ is the observer's internal state
initialized at $\hat{z}(0)$, which processes the measured output and control
input from~\eqref{apd:eq_system} to provide an estimate
$\hat{x}(t) \in \mathbb{R}^{n_{x}}$. Designing the observer requires
choosing the continuous functions
$\Phi: \mathbb{R}^{n_{z}} \times \mathbb{R}^{n_{y}} \times
\mathbb{R}^{m} \rightarrow \mathbb{R}^{n_{z}}$ and
$\Psi: \mathbb{R}^{n_{z}} \times \mathbb{R}^{n_{y}} \rightarrow
\mathbb{R}^{n_{x}}$ such that the estimation error:
\begin{equation}
\xi(t) \mathrel{:=} x(t) - \hat{x}(t)
\label{apd:eq_estimation_error}
\end{equation}
globally asymptotically converges to zero as $t \rightarrow \infty$, i.e.,
$\forall x_{0} \in \mathcal{X}, \hat{z}(0) \in \mathbb{R}^{n_{z}}
\colon \lim_{t \to \infty} \|\xi(t)\|= 0$.

\subsection{KKL Observer Theory for Autonomous Systems}\label{apd:KKL_Observer_Theory_for_Autonomous_Systems}

Originally established for autonomous systems ($\dot{x} = f(x)$), the KKL
theory states the existence of an immersion mapping $\mathcal{T}$. According
to~\citet{brivadis2023further}, if the system is \textit{backward
distinguishable}\footnote{A system is backward
$\mathcal{O}$-distinguishable on $\mathcal{X}$ if for every distinct pair
$x_0^1, x_0^2 \in \mathcal{X}$, there exists $\tau < 0$ such that the
backward solutions satisfy
$h(x(\tau; x_0^1)) \neq h(x(\tau; x_0^2))$.} and \textit{forward complete}
on a compact set $\mathcal{X}$, there exists an injective transformation
$\mathcal{T}: \mathcal{X} \to \mathbb{R}^{n_z}$ that maps the nonlinear
dynamics into a linear latent system:
\begin{equation}
\dot{z}(t) = A z(t) + B h(x(t)), \quad z(0) = \mathcal{T}(x_0)
\label{apd:eq_autonomous_latent}
\end{equation}
where $A \in \mathbb{R}^{n_z \times n_z}$ is a Hurwitz matrix and
$B \in \mathbb{R}^{n_z \times n_y}$ is an output injection term, and the
pair $(A, B)$ is chosen such that it is controllable\footnote{The pair
$(A, B)$ is controllable if
$\text{rank} [B \ AB \ \cdots \ A^{n_z-1}B] = n_z$.} and the latent
dimension satisfies $n_z = n_y(2n_x + 1)$. The transformation
$\mathcal{T}(x)$ is the solution to the
PDE~\eqref{eq:KKL_autonomous_PDE}, restated here:
\begin{equation}
\frac{\partial \mathcal{T}}{\partial x}(x)\, f(x)
= A\,\mathcal{T}(x) + B\,h(x), \quad \mathcal{T}(0) = 0
\label{apd:eq_autonomous_PDE}
\end{equation}
Once $\mathcal{T}(x)$ is identified, a state estimate $\hat{x}$ can be
recovered using the left-inverse $\mathcal{T}^*$ by simulating the linear
observer~\eqref{eq:KKL_observer_latent}:
\begin{equation}
\begin{cases}
\dot{\hat{z}}(t) = A\hat{z}(t) + B\,y(t), \\
\hat{x}(t) = \mathcal{T}^*(\hat{z}(t))
\end{cases}
\label{apd:eq_autonomous_observer}
\end{equation}
Since $A$ is Hurwitz, the error $e(t) = z(t) - \hat{z}(t)$ converges
asymptotically to zero~\citep[Remark 4]{niazi2025kkl}.

\subsection{Extension to Non-Autonomous Systems}\label{apd:KKL_Observer_Extension}

Extending the framework to non-autonomous systems (i.e., systems with inputs
$u$) introduces a significant constraint, as the transformation
$\mathcal{T}(x)$ must satisfy the immersion condition uniformly for all
admissible inputs $u$. Following the physics-informed learning approach
introduced in~\citet{niazi2025kkl}, we seek a transformation $\mathcal{T}$
that maps the nonlinear dynamics to a linear latent system. To ensure robust
estimation, this target linear system is designed to be Bounded-Input
Bounded-State (BIBS) stable. The target dynamics in the latent space are
formulated as:
\begin{equation}
\dot{z} = A\,z(t) + B\,h(x(t)) + \bar{\varphi}(u(t))
\label{apd:eq_nonautonomous_latent}
\end{equation}
where $\bar{\varphi}: \mathbb{R}^{m} \to \mathbb{R}^{n_z}$ is a design
parameter of choice (typically linear) representing the input injection.
In the stationary approach of
Section~\ref{subsection:KKL_non_autonomous}, this corresponds to the
injection term $\bar{\varphi}(\hat{z})\,u$. Consequently, the governing PDE
for the transformation becomes:
\begin{equation}
\frac{\partial \mathcal{T}}{\partial x}\, f(x, u)
= A\,\mathcal{T}(x) + B\,h(x) + \bar{\varphi}(u)
\label{apd:eq_nonautonomous_PDE}
\end{equation}
Solving the PDE in~\eqref{apd:eq_nonautonomous_PDE} analytically is often
intractable. Recent work by~\citet{niazi2025kkl} utilizes PINNs to
approximate $\mathcal{T}(x)$ by minimizing the residual of the PDE
directly, effectively turning the observer design problem into a learning
problem. The Dynamic HyperKKL framework
(Section~\ref{subsubsec:dynamic}) extends this further by allowing
$\mathcal{T}$ itself to vary with the input history via the time-dependent
PDE~\eqref{eq:dynamic_pde}.

\clearpage
\section{Simulation Results}
\label{apd:sim_results}


\begin{figure}[h!]
    \centering
    \begin{subfigure}[b]{0.48\textwidth}
        \centering
        \includegraphics[width=\textwidth]{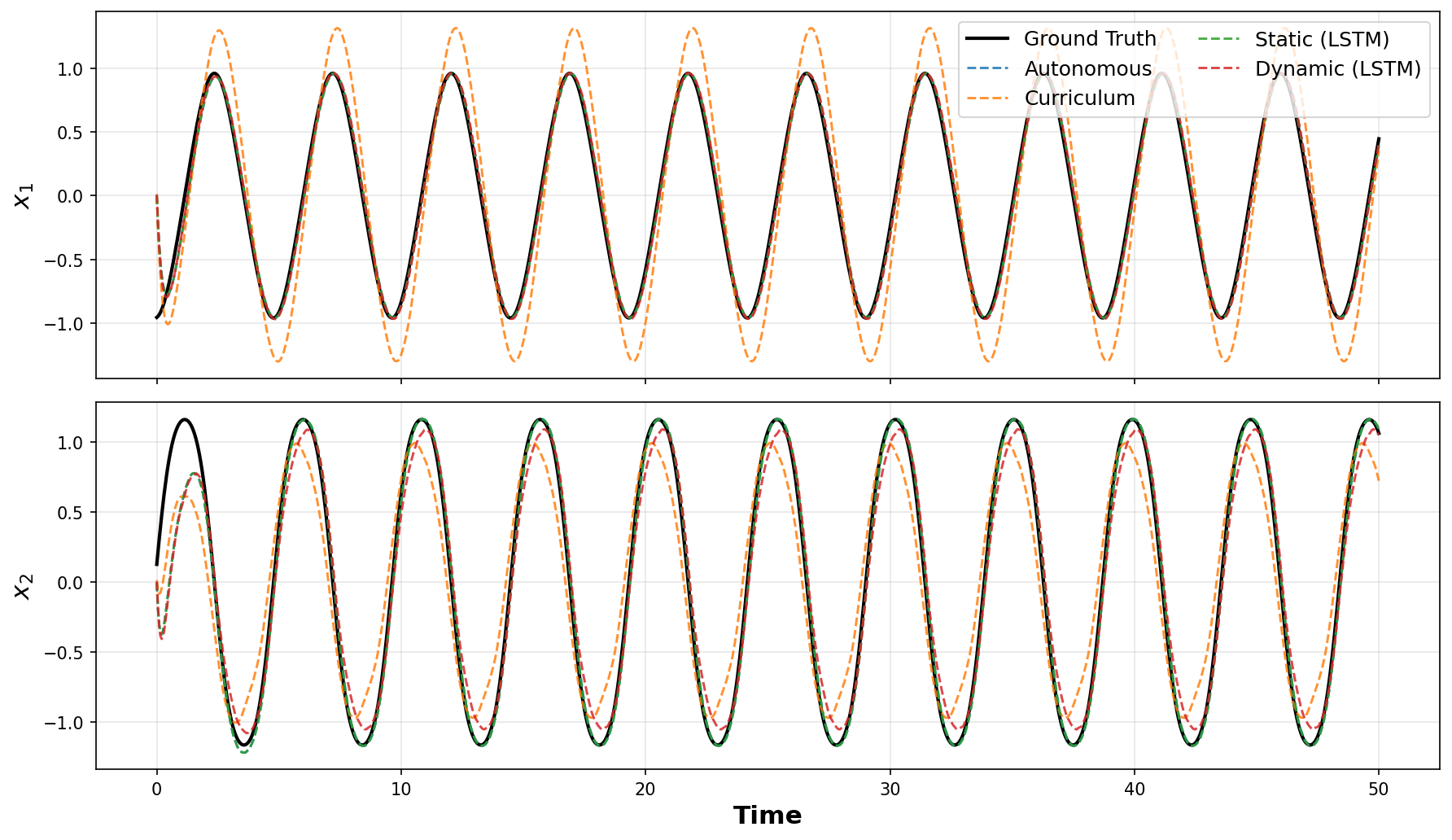}
        \caption{Zero Input}
    \end{subfigure}
    \hfill
    \begin{subfigure}[b]{0.48\textwidth}
        \centering
        \includegraphics[width=\textwidth]{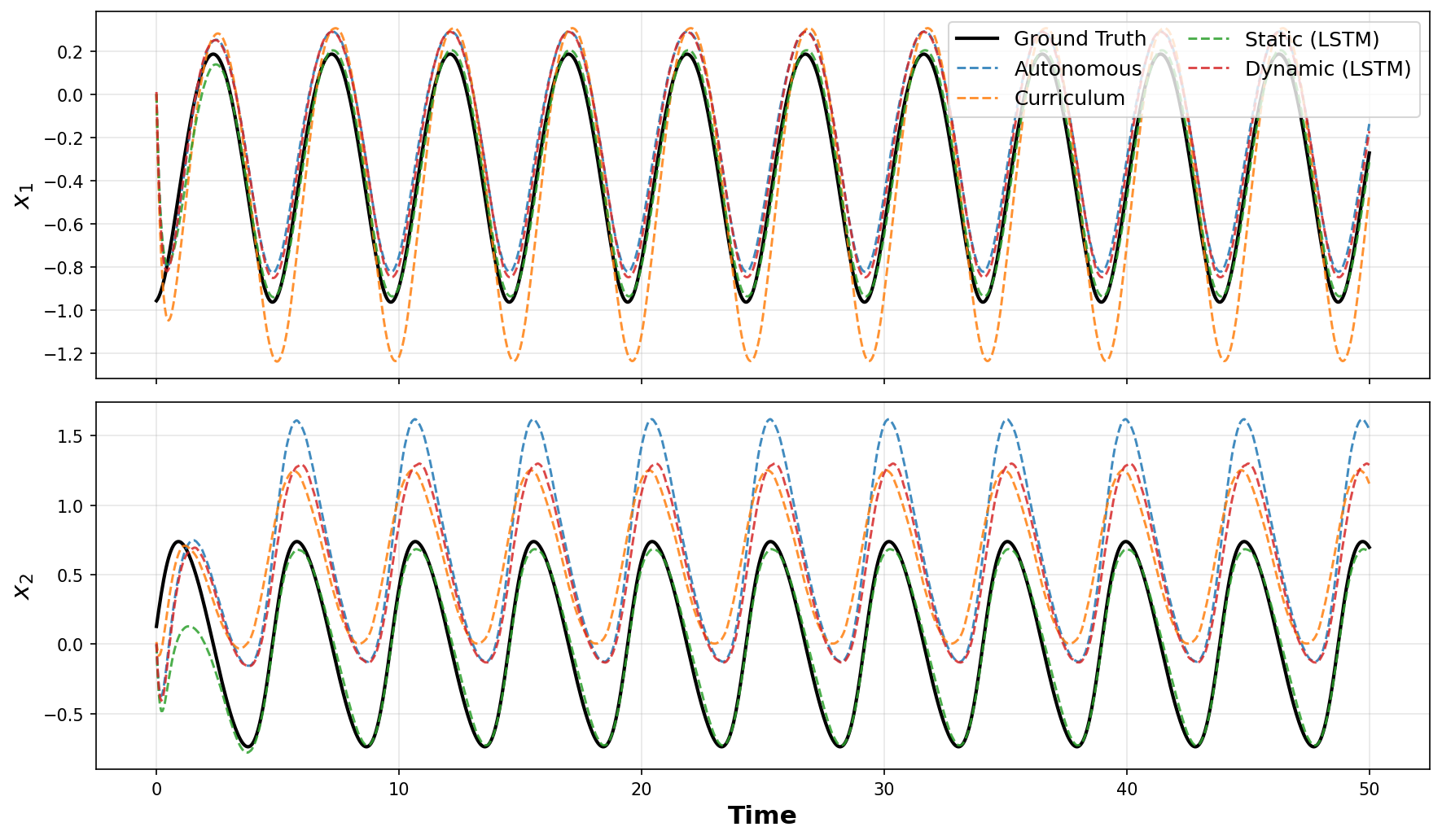}
        \caption{Constant Input}
    \end{subfigure}
    
    \vspace{0.5em} 
    
    \begin{subfigure}[b]{0.48\textwidth}
        \centering
        \includegraphics[width=\textwidth]{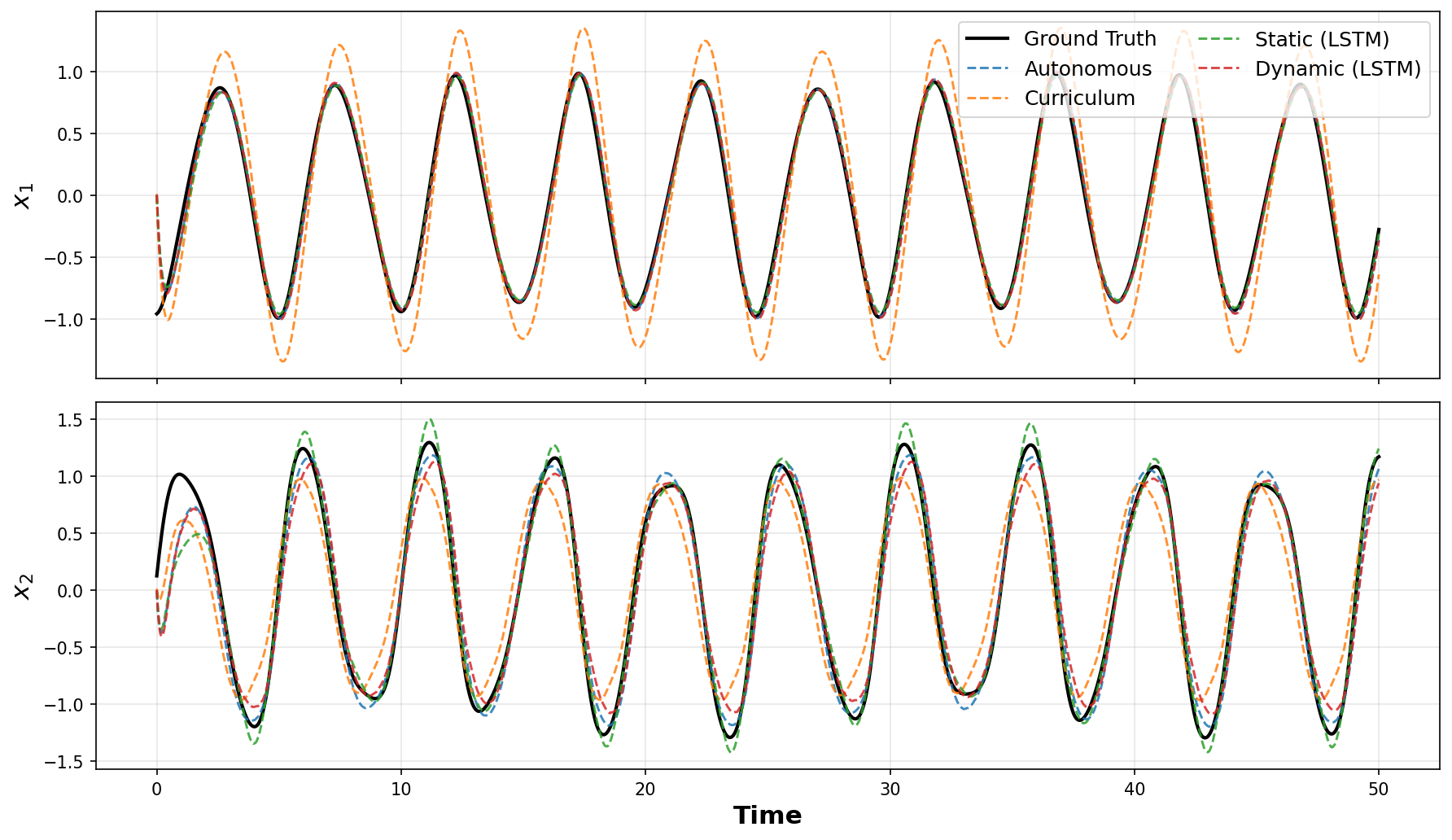}
        \caption{Sinusoid Input}
    \end{subfigure}
    \hfill
    \begin{subfigure}[b]{0.48\textwidth}
        \centering
        \includegraphics[width=\textwidth]{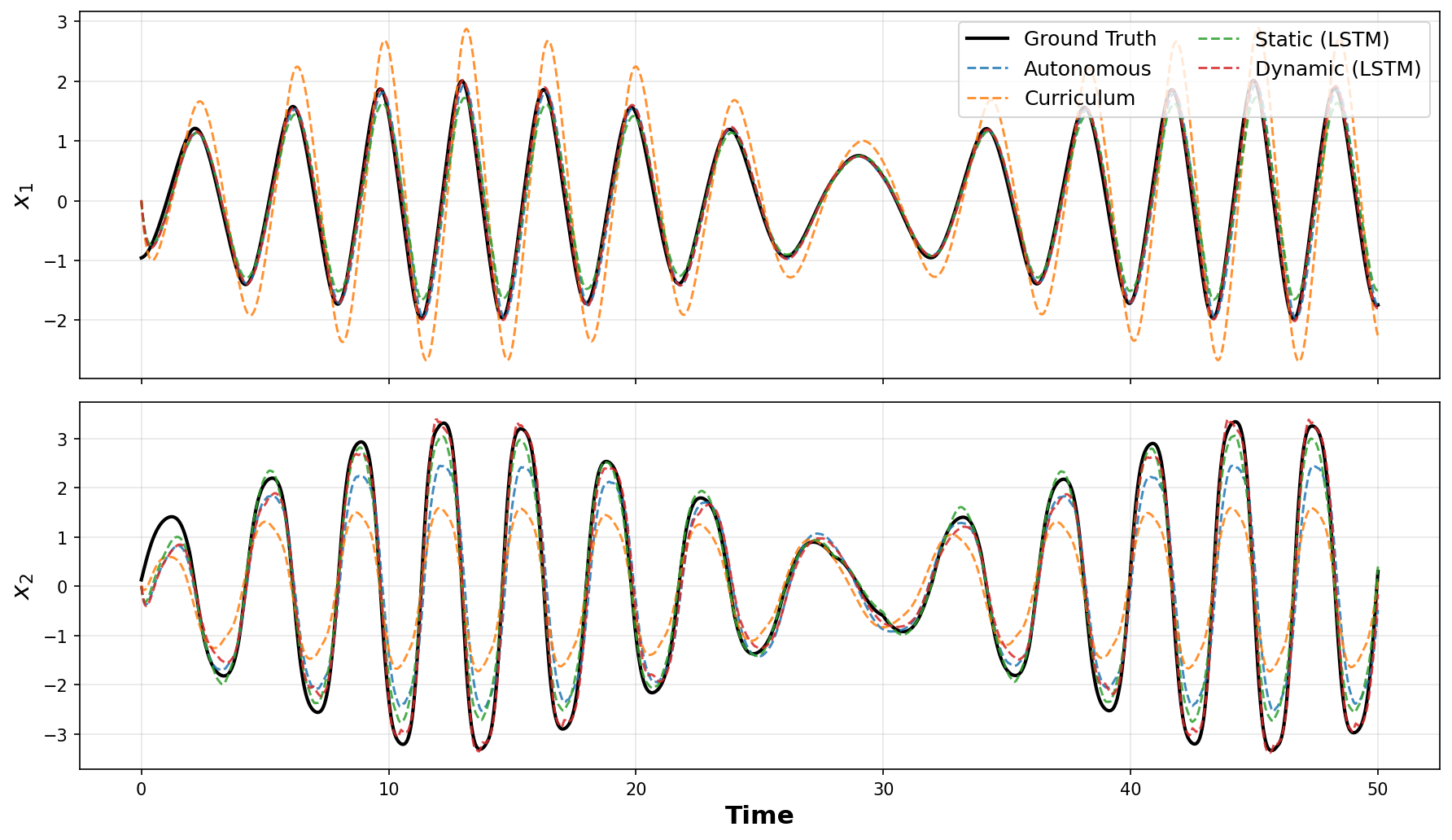}
        \caption{Square Input}
    \end{subfigure}
    \caption{Duffing System: State estimation time-series.}
    \label{fig:duffing_results}
\end{figure}

\begin{figure}[h!]
    \centering
    \begin{subfigure}[b]{0.48\textwidth}
        \centering
        \includegraphics[width=\textwidth]{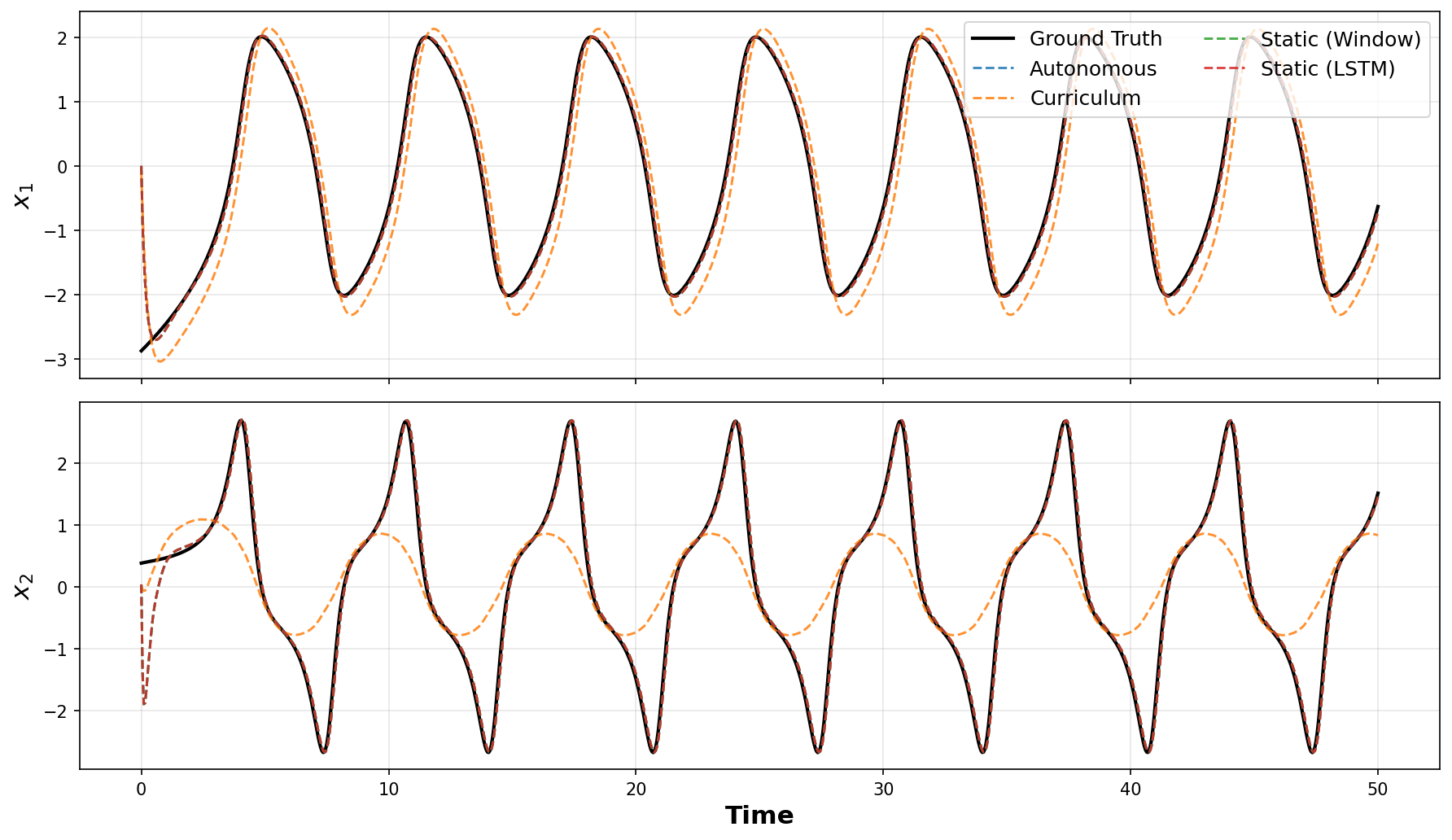}
        \caption{Zero Input}
    \end{subfigure}
    \hfill
    \begin{subfigure}[b]{0.48\textwidth}
        \centering
        \includegraphics[width=\textwidth]{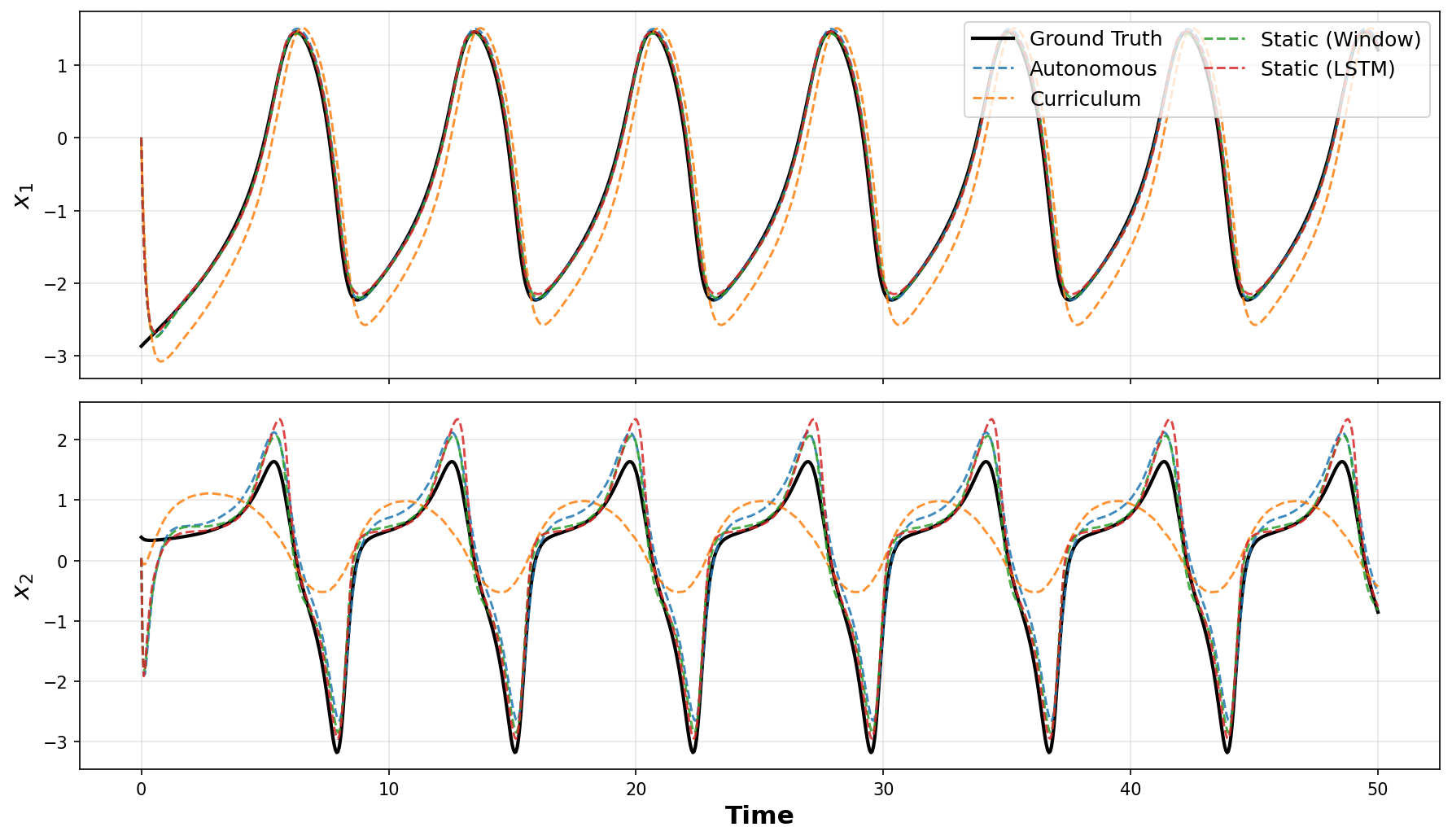}
        \caption{Constant Input}
    \end{subfigure}
    
    \vspace{0.5em}
    
    \begin{subfigure}[b]{0.48\textwidth}
        \centering
        \includegraphics[width=\textwidth]{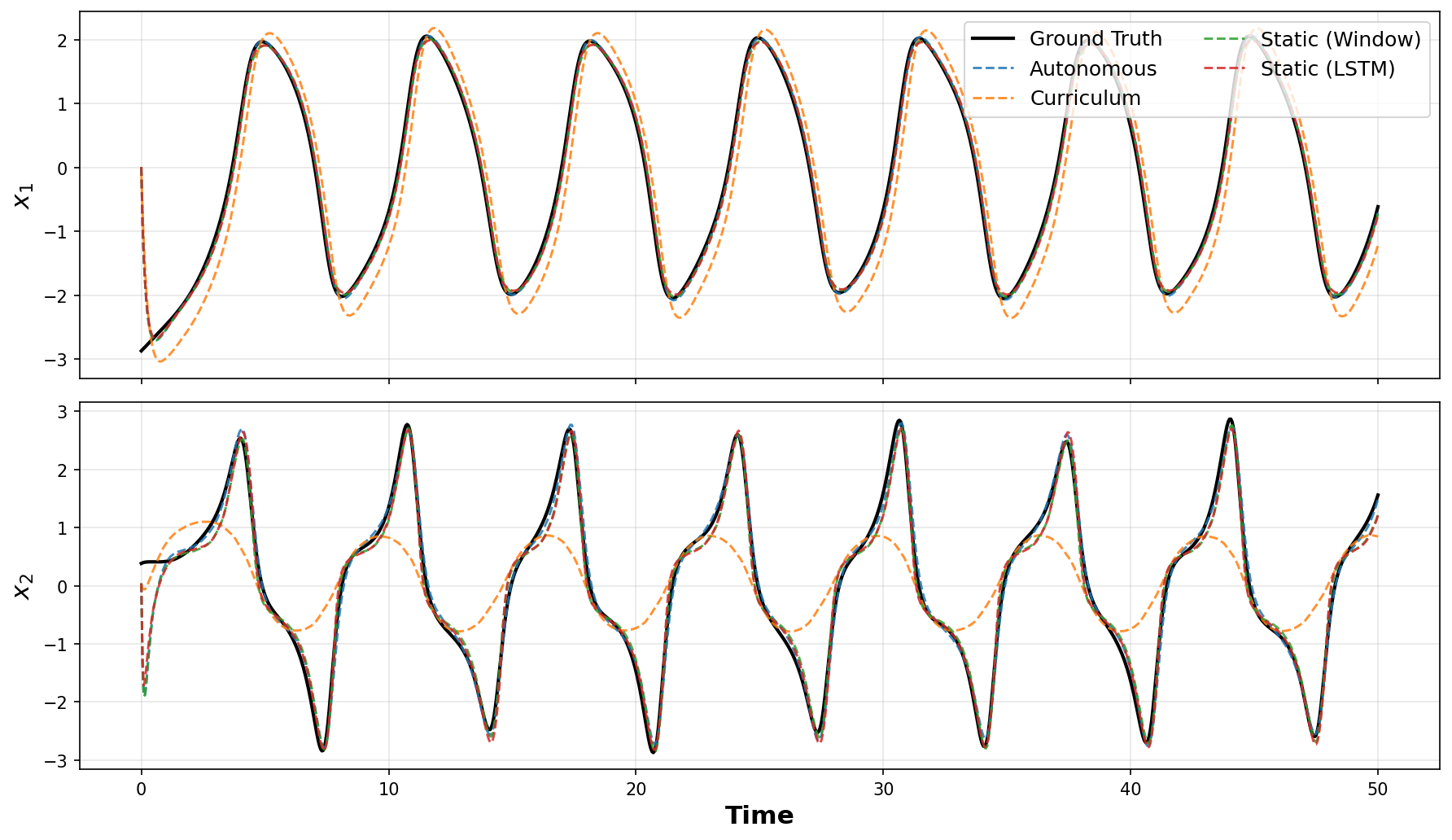}
        \caption{Sinusoid Input}
    \end{subfigure}
    \hfill
    \begin{subfigure}[b]{0.48\textwidth}
        \centering
        \includegraphics[width=\textwidth]{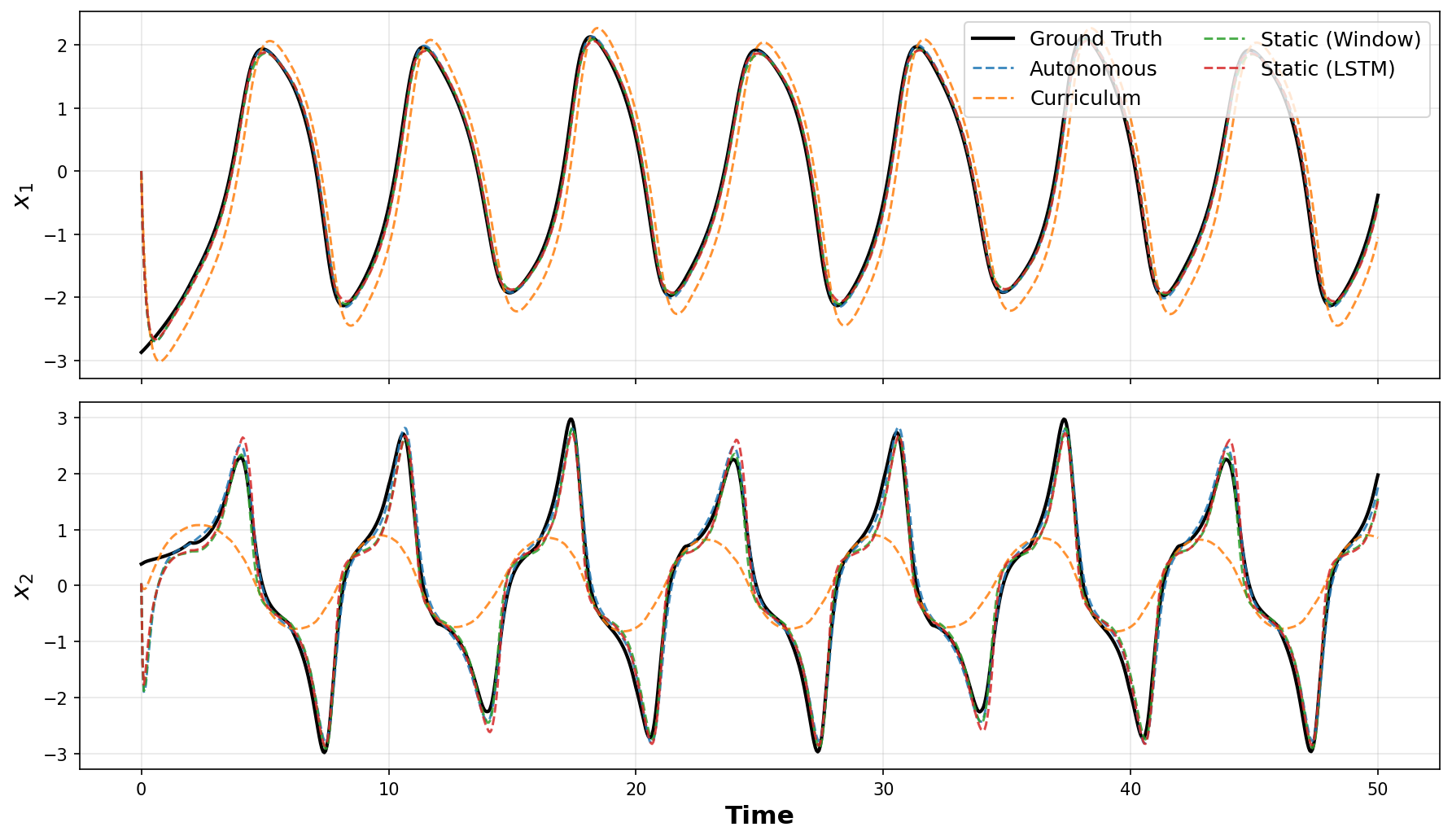}
        \caption{Square Input}
    \end{subfigure}
    \caption{Van der Pol System: State estimation time-series.}
    \label{fig:vdp_results}
\end{figure}

\clearpage 
\begin{figure}[h!]
    \centering

    \includegraphics[width=0.48\textwidth]{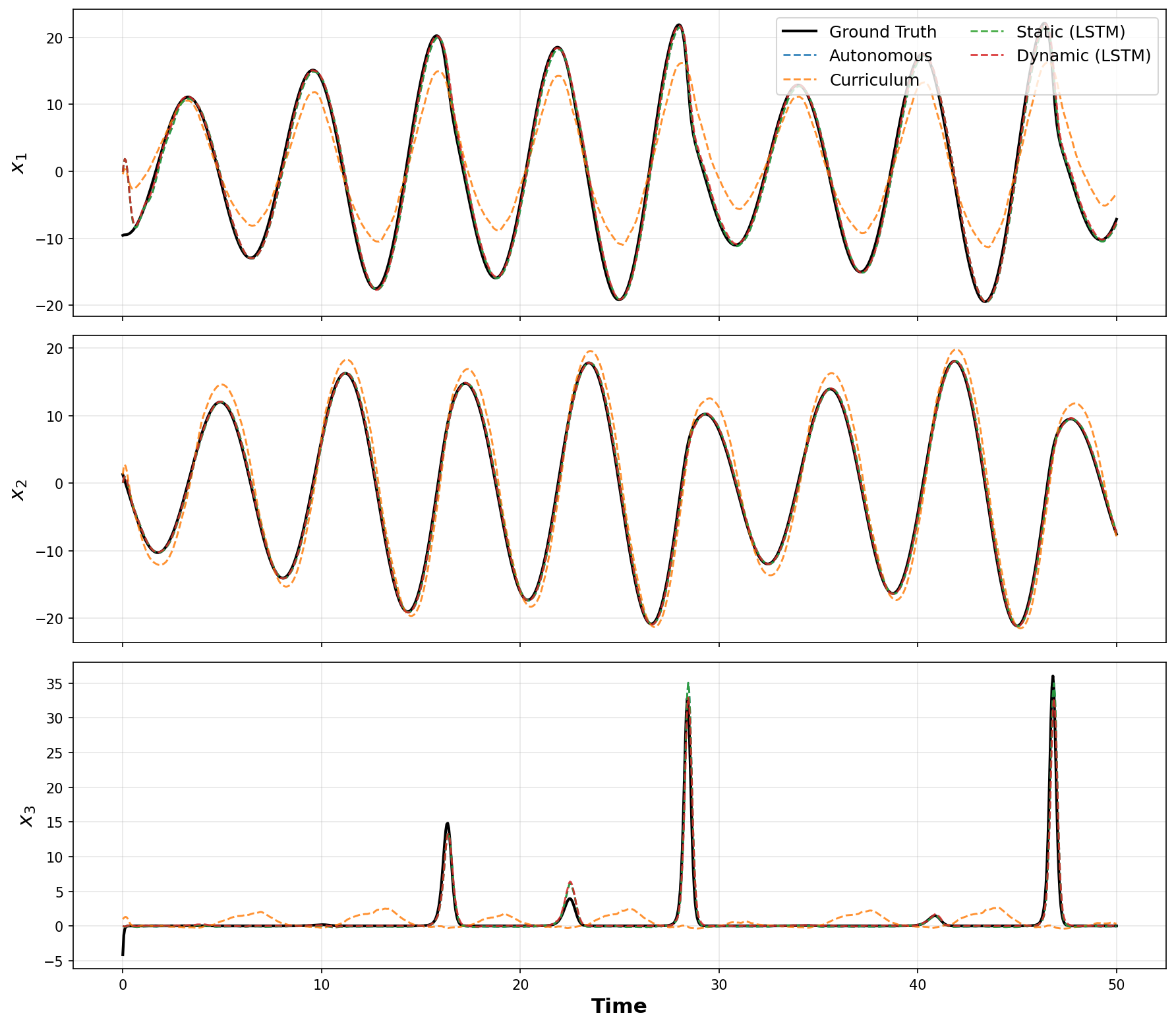}
    \hfill
    \includegraphics[width=0.48\textwidth]{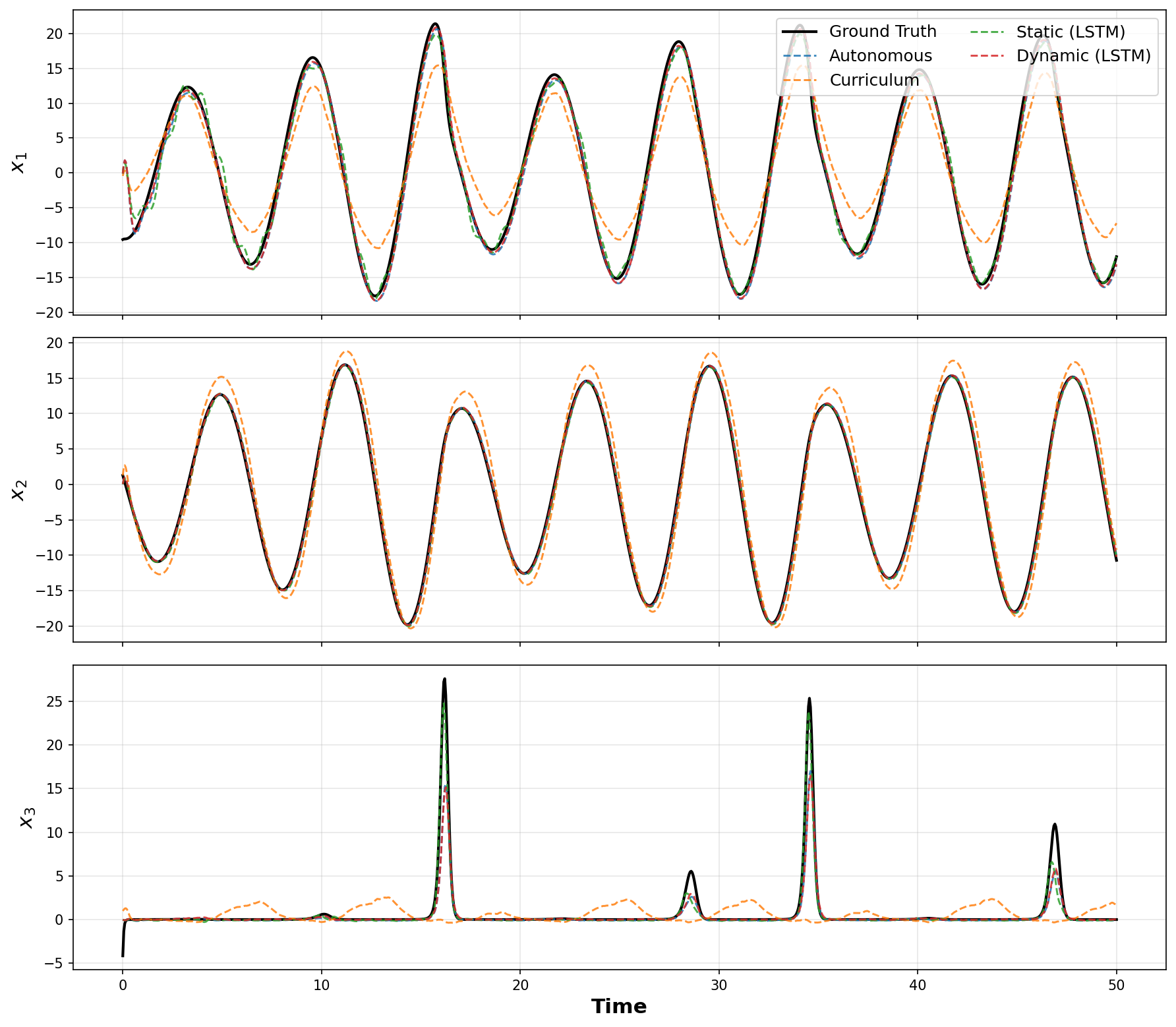}

    \vspace{0.5em}

    \includegraphics[width=0.48\textwidth]{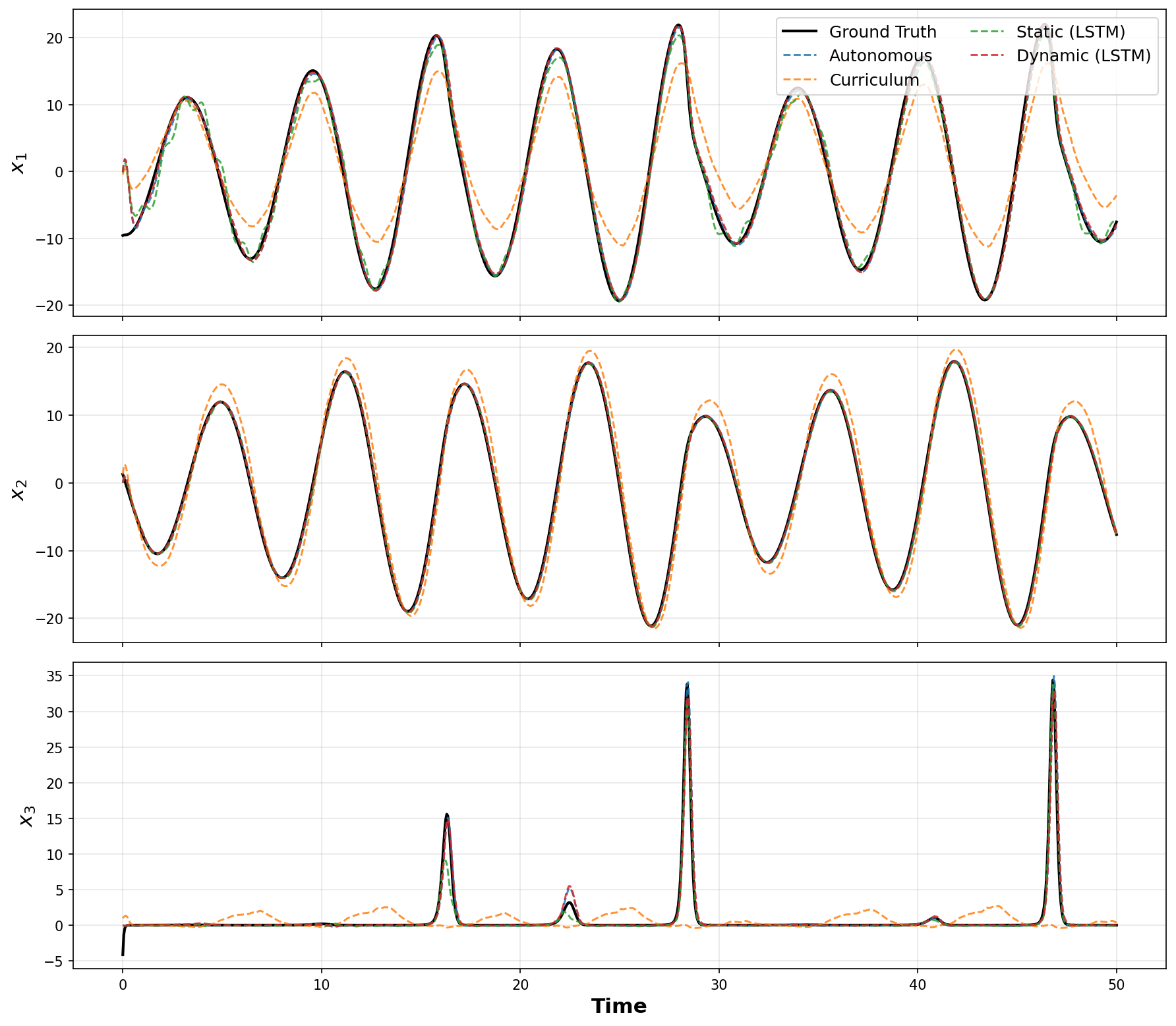}
    \hfill
    \includegraphics[width=0.48\textwidth]{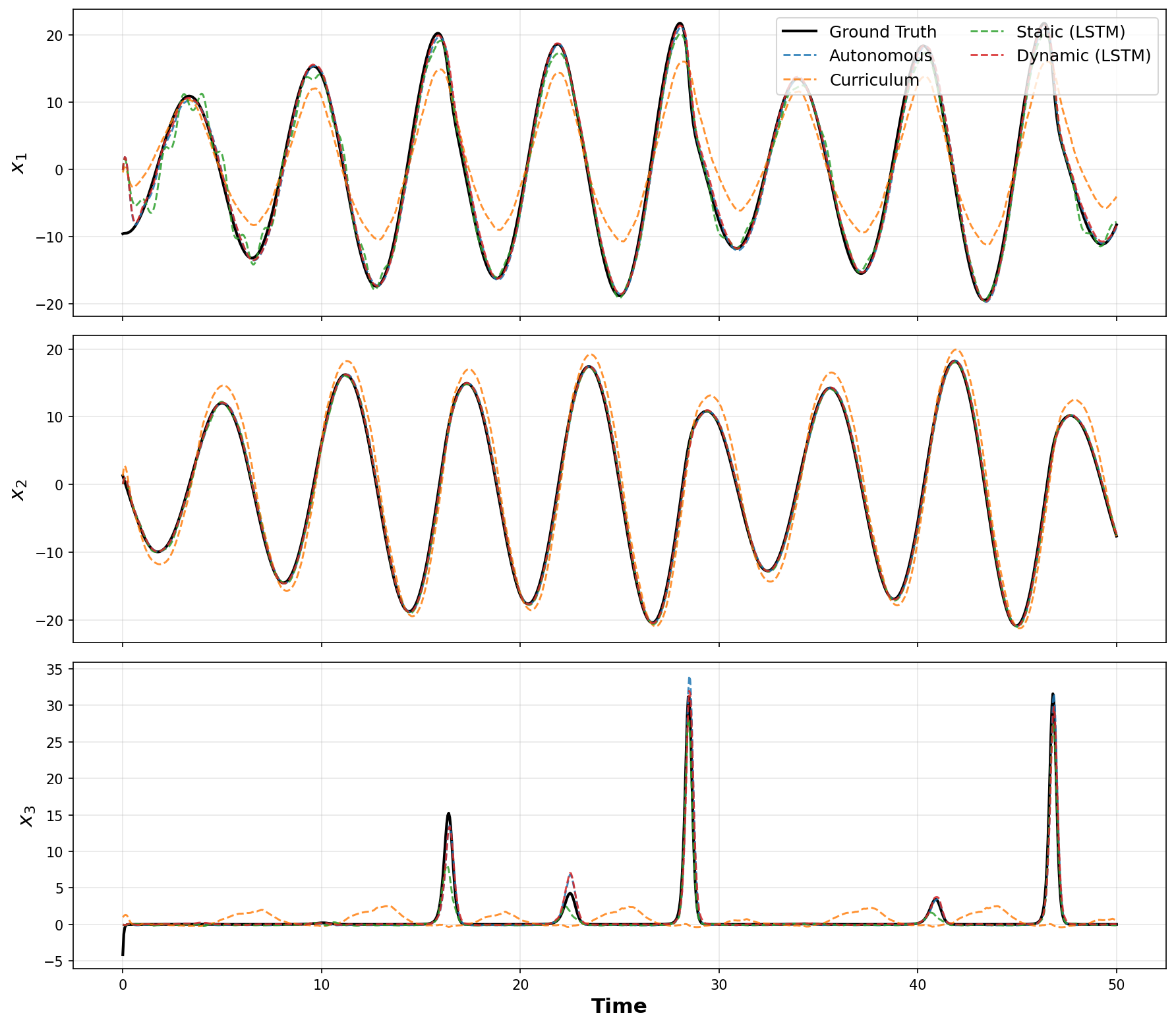}

    \caption{Rossler System (Chaotic): State estimation time-series.}
    \label{fig:rossler_results}
\end{figure}

\begin{figure}[h!]
    \centering

    \includegraphics[width=0.48\textwidth]{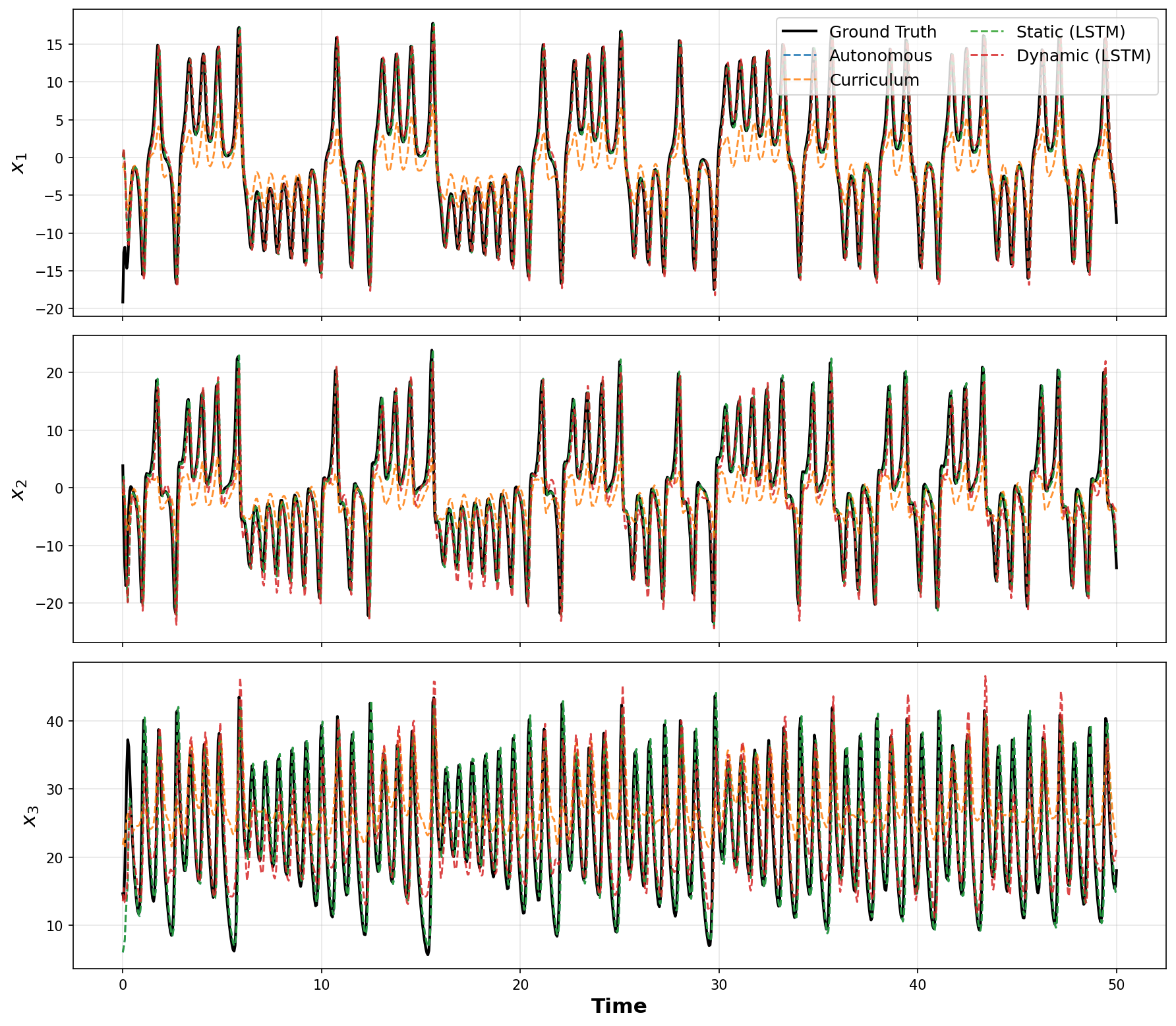}
    \hfill
    \includegraphics[width=0.48\textwidth]{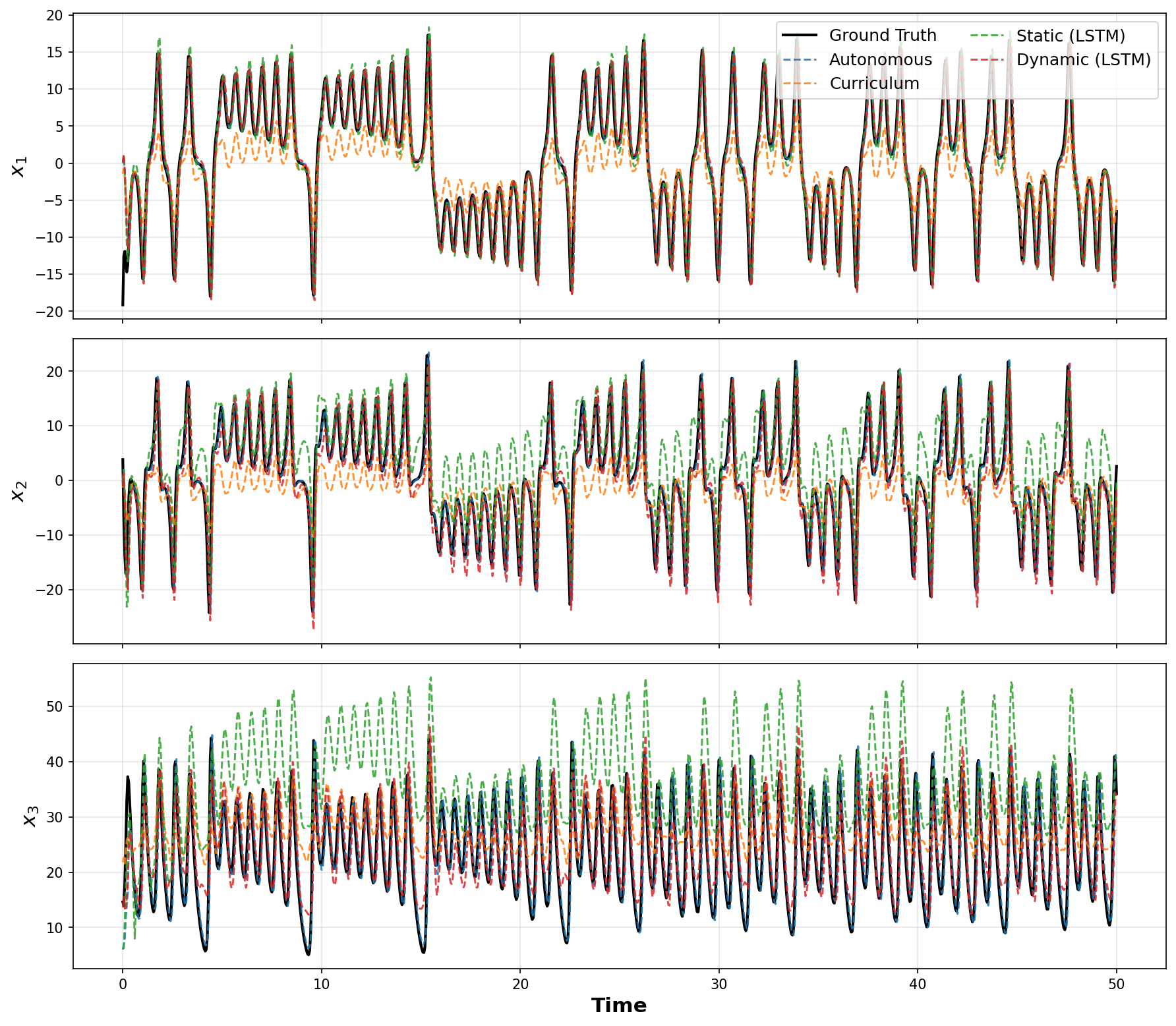}

    \vspace{0.5em}

    \includegraphics[width=0.48\textwidth]{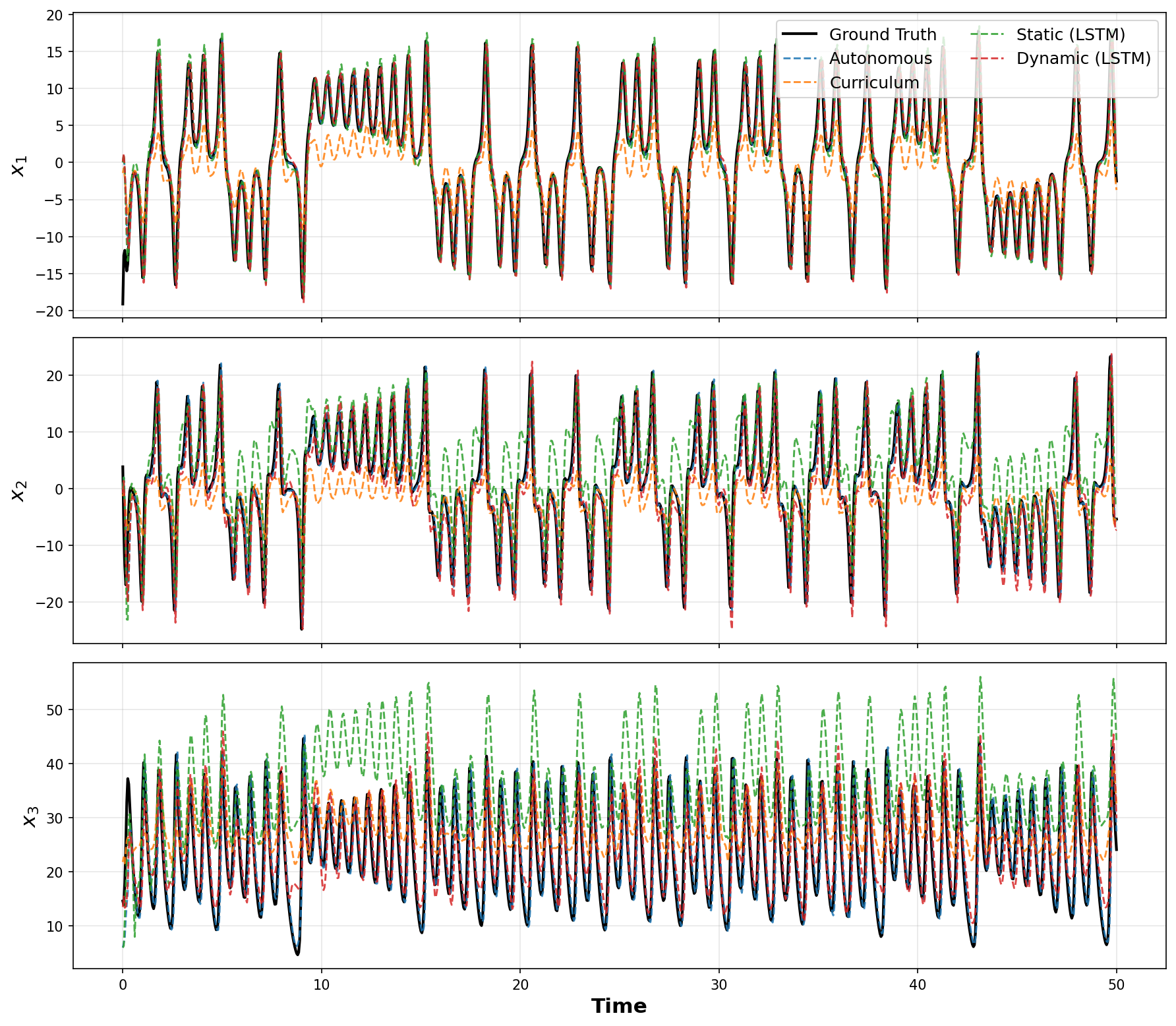}
    \hfill
    \includegraphics[width=0.48\textwidth]{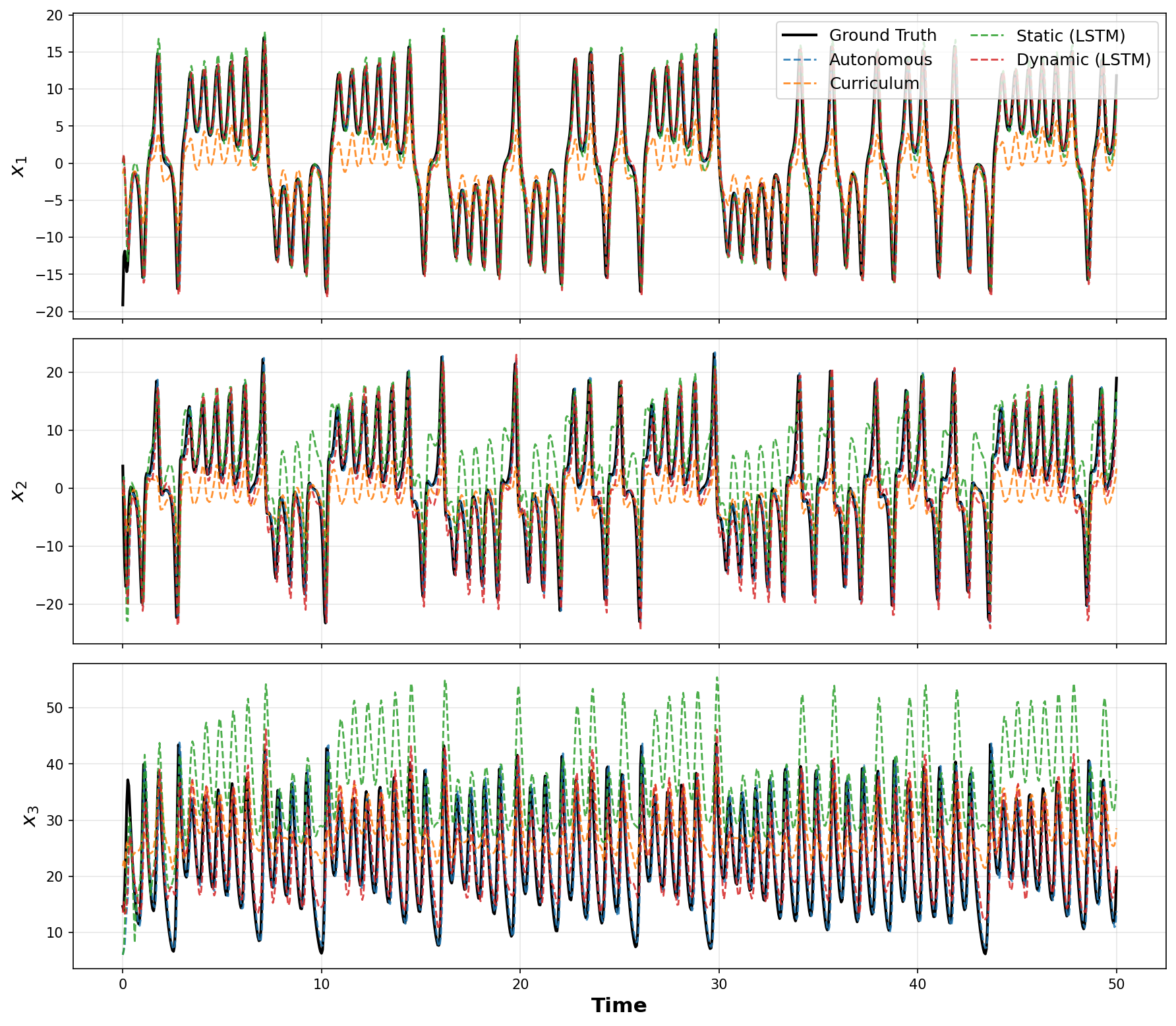}

    \caption{Lorenz System (Chaotic): State estimation time-series.}
    \label{fig:lorenz_results}
\end{figure}


\end{document}